\documentclass[useAMS,usenatbib]{mn2e}
\usepackage{epsfig,graphicx,lscape}
\usepackage{amsmath}
\newcommand{\kms}{{\,km\,s}$^{-1}$}

\usepackage{rotating}
\usepackage{float,longtable}
\usepackage{url}

\title[Time-variable interstellar absorption]{Early-type stars observed in the 
ESO UVES Paranal Observatory Project - V. Time-variable interstellar absorption}

\author[C. M. McEvoy et al.]{Catherine M. McEvoy$^{1}$\thanks{email: {cmcevoy14@qub.ac.uk}},
Jonathan~V. Smoker$^{2}$, Philip~L. Dufton$^{1}$, \newauthor
Keith T. Smith$^{2, 3}$, Michael B. Kennedy$^{1}$,  Francis~P. Keenan$^{1}$,\newauthor
David L. Lambert$^{4}$, Daniel E. Welty$^{5}$, James~T. Lauroesch$^{6}$
\\
$^{1}$
Astrophysics Research Centre, 
School of Mathematics and Physics, 
Queen's University Belfast, 
Belfast BT7 1NN, UK
\\
$^{2}$
European Southern Observatory, 
Alonso de Cordova 3107, 
Casilla 19001, 
Vitacura, 
Santiago 19, Chile
\\
$^{3}$
Royal Astronomical Society
Burlington House,
Piccadilly,
London, W1J 0BQ,
UK
\\
$^{4}$ Department of Astronomy,
The University of Texas at Austin,
RLM 16.316,
Austin, TX 78712
\\
$^{5}$ Astronomy \& Astrophysics Center,
University of Chicago,
5640 South Ellis Ave.,
Chicago, IL 60637, USA
\\
$^{6}$ Department of Physics and Astronomy, 
University of Louisville, 
Louisville, 
KY 40292, 
USA
}

\date{Accepted 2015 April 27. Received 2015 April 23; in original form 2015 January 26}

\pubyear{2014}

\begin{document}
\label{firstpage}
\maketitle

\begin{abstract}
The structure and properties of the diffuse interstellar medium (ISM) on small scales, sub-au to 1 pc, are poorly understood. We compare interstellar absorption-lines, 
observed towards a selection of O- and B-type stars at two or more epochs, to search for variations over time caused by the transverse motion of each star combined with 
changes in the structure in the foreground ISM. Two sets of data were used: 83 VLT-UVES spectra with approximately 6 yr between epochs and 21 McDonald observatory 2.7m 
telescope echelle spectra with 6 -- 20 yr between epochs, over a range of scales from $\sim$ 0 -- 360 au. The interstellar absorption-lines observed at the two epochs were
subtracted and searched for any residuals due to changes in the foreground ISM. Of the 104 sightlines investigated 
with typically five or more components in Na\,{\sc i} D, possible temporal variation was identified 
in five UVES spectra (six components), in Ca\,{\sc ii}, Ca {\sc i} and/or Na\,{\sc i} absorption-lines. The variations detected range from 7\% to a factor of 3.6 in column density. No variation was 
found in any other interstellar species. Most sightlines show no variation, with 3$\sigma$ upper limits to changes of the order 0.1 -- 0.3 dex in 
Ca\,{\sc ii} and Na\,{\sc i}. These variations observed imply that fine-scale structure is present in the ISM, but at the resolution available in this study, is not very common 
at visible wavelengths. A determination of the electron densities and lower limits to the total number density of a sample of the sightlines implies that there is no 
striking difference between these parameters in sightlines with, and sightlines without, varying components.
\end{abstract}


%
\section{Introduction}                                         
\label{s_intro}

The properties of the interstellar medium (ISM) on small scales, from sub-astronomical unit (au) to 1 parsec, remain poorly understood. The first indication that au-scale structures were present in H\,{\sc i} was provided by \citet{dia89}, who used Very Long Baseline Interferometry (VLBI) radio observations to identify changes in the 21~cm absorption towards three extragalactic radio sources as a function of resolution. This suggested foreground structures with linear dimensions of only $\sim$25~au. Further absorption-line VLBI measurements in H\,{\sc i} \citep{fai98,bro05,laz09,roy12} and molecules \citep{mar93,lis00} also found tiny-scale structures, although with relatively small plane-of-sky covering fractions of $\sim$10 per cent. Similar results have been seen via multi-epoch H\,{\sc i} absorption-line measurements towards pulsars \citep[e.g.][]{Fra94,joh03,sta10} and indicate that, in H\,{\sc i}, tiny-scale variations are a common phenomenon but with a low probability of intercepting any given line of sight.

Optical and UV observations have also been used to detect temporal and spatial variation in interstellar absorption. Most have relied upon multi-epoch ultra-high resolution optical observations in Na\,{\sc i} and Ca\,{\sc ii} \citep[e.g][]{hob91,bla97,pri00,mey12}, although some authors have also detected variations in K\,{\sc i}, Ca\,{\sc i} and/or Fe\,{\sc i} \citep{cra00,smi13,gal13}. Binary systems \citep[e.g.][]{mey90,cor13} and star clusters \citep[e.g.][]{and01,van09} have also been used to study the fine-scale structure of the ISM. \citet{wat96} found differences in Na {\sc i} towards the individual members of all 17 of the binary or multiple star systems they observed, with separations 480-29000 au. These observations imply a population of dense au-scale clouds in the ISM or changes in ionization or depletion over time, but are restricted to trace neutral species. \cite{boi13} studied the spatial distribution of CH$^+$, CH and CN on scales ranging 1--20 au for a single sightline towards Zeta Per, finding the neutral species to be uniformly distributed and a CH$^+$ excess implying its production in very small localized active regions of only a few au.  Data for UV absorption-lines can probe dominant ions and provide further information on the physical conditions; time-variable UV lines have been found by e.g. \citet{dan01} and \citet{wel07}. Further examples of time-variable interstellar lines are cited in the reviews by \citet{cra03} and \citet{lau07}.

We note it is difficult to decipher between apparent spatial and temporal variations in the foreground ISM.
The main questions concerning these variations are as follows. What are the physical conditions in these structures? What is their formation mechanism? How stable are they and how do the structures seen in the optical relate to the H\,{\sc i} absorption-line measurements?

This is one of a series of papers which use data on early-type stars in the UVES Paranal Observatory Project \citep[POP;][]{bag03} \footnote{\url{http://www.eso.org/uvespop}} to study the ISM.
In \citet[Paper I]{hun06} the near-UV interstellar lines of Na\,{\sc i}, Ti\,{\sc ii} and Ca\,{\sc ii} were observed towards 74 O- and B-type stars, and basic correlations in the observed column densities were investigated.
\citet[Paper II]{smo07} combined these data with H\,{\sc i} 21~cm observations to estimate distances to intermediate- and high-velocity clouds, while
\citet[Paper III]{smo11} used POP observations of stars in open clusters (IC\,2391, NGC\,6475 and Praesepe) combined with archival multi-epoch data to search for small- (pc) and tiny- (au) scale structure of the 
ISM. 
They searched for variations in Ti\,{\sc ii}, Ca\,{\sc ii} K, Na\,{\sc i} D and K\,{\sc i} absorption; a total of three possible examples of tiny-scale structure were found.

Here we present UVES observations at additional epochs of the same POP O- and B-type lines of sight, coupled with 
additional data taken at McDonald observatory to search for time-variable absorption. The time between the two epochs was $\sim$6~yr for the UVES POP sample and $\sim$ 6 -- 20~yr for the McDonald data. Due to the proper motion of the stars, it is hence possible to study changes in their lines of sight on scales of $\sim$ 0 -- 360 au. We aim to extend the sample size of twin epoch observations to determine just how ubiquitous the small scale structure of the ISM is. To our knowledge this is the largest twin-epoch spectroscopic survey to date.

This paper is organized as follows. Section~\ref{s_obs} describes the archival data used, the new observations, and the data reduction, while Section~\ref{s_results} contains the main results. In Section~\ref{s_dis} we discuss the upper limits and possible time-varying interstellar line strength in terms of different models of the ISM. Finally, Section~\ref{s_summary} presents a summary of our results and suggests avenues for future research. 

\section{Observations, archival data, and data reduction}                                        
\label{s_obs}

\subsection{UVES data}
Our study uses a combination of archival and new observations. Archival data from the ESO POP (ESO programme ID 266.D-5655(A), \citealt{bag03}) were used to provide a catalogue of high resolution ($R\equiv\frac{\lambda}{\mathrm{d}\lambda}\sim 80,000$), covering the wavelength range 304 -- 1040 nm for about 400 stars obtained with the upgraded UVES spectrometer \citep{Dek00, Smo09} of the ESO Very Large Telescope. The signal-to-noise ratio (S/N) of the stellar spectra is high, ranging 200 -- 800, and averaging 520 around Na {\sc i}, and ranging 100 -- 800, averaging 330 around Ca {\sc ii}. The POP survey contains a total of 98 O- and B-type stars, suitable for use as background sources to study the ISM. These sightlines form the first epoch of data for most stars in our sample, and were observed between 2001 and 2003. The merged versions of the spectra were taken from the POP website (\citep[\url {http://www.eso.org/uvespop}][]{bag03}). 

New second-epoch observations of the same O- and B-type stars were taken during 2008 (ESO programme ID 081.C-0475(A)) 
on the same UVES instrument to allow direct comparison between the two epochs at the same spectral 
resolution. It is important to note that the measured resolution of the UVES data can change by up to 10 \% over several yr, depending on the season \footnote{\url{http://www.eso.org/observing/dfo/quality/UVES/reports/HEALTH/trend_report_ECH_RESOLUTION_DHC_HC.html}}. We have investigated the effects of these small resolution changes on the column densities calculated and found that in the majority of cases, the differences are within the error estimates (see Sect. \ref{s_columndensity}). The (390+580) setting was used giving wavelength coverage from 328--450 nm and 475--681 nm, with 
a gap from 578--581 nm where the two CCDs in the red mosaic are joined. For some sightlines where changes were detected, we obtained third epoch data in 2013 using the same instrumental setup (ESO programme ID 092.C-0173(A)). All of these data were reduced using the ESO UVES pipeline \citep{bal00} and the multiple exposures at each epoch co-added using \textsc{iraf}\footnote{IRAF is distributed by the National Optical Astronomy Observatory, which is operated by the Association of Universities for Research in Astronomy (auRA) under cooperative agreement with the National Science Foundation.}, to bring the epoch 2 and 3 data to the same point of reduction as epoch 1. As the epoch 1 data were taken from the POP website, they had been reduced using an earlier version of the pipeline than the epoch 2 and 3 data. To investigate the effects of the different pipelines on the spectral resolution, a subset of epoch 1 data (including some showing possible variation) were re-reduced using a more recent pipeline. No significant resolution changes were discovered and so the merged version of the data from the POP website was deemed sufficient for comparison.

From this stage onwards all sets of UVES data were treated in exactly the same manner. Each spectrum was
analysed in {\sc idl}. Interstellar absorption-lines of interest (listed in Table 
\ref{t_Species}) were isolated and each line from epoch 1 was cross correlated with its equivalent from epoch 2 or 3 to correct for any velocity 
shifts between the ISM lines in the two epochs, and to bring both epochs into kinematic local-standrad-of-rest frame. The two epochs were then overlaid so that any obvious changes in the line profiles could be identified by 
eye. Spectra were renormalized to remove the 
stellar continua and the Na\,{\sc i} D spectra corrected for telluric absorption features where necessary, by dividing by a scaled telluric 
spectrum created from our set of observations  with regions free of Na absorption or from Cerro Paranal Sky Model {\sc skycalc}
\footnote{http://www.eso.org/observing/etc/} 
\citep{nol12, jon13}. In all cases where possible variations were identified,they are seen in both the Na D$_1$ and Na D$_2$ lines, which are affected differently by telluric contamination, and so inconsistent correction of telluric lines could not account for the variation seen. epoch 1 spectra were subtracted from epoch 2 or 3 for each absorption feature to search for any residuals that would imply variations in the 
line of sight between epochs. It was also necessary to compare these residual plots to residuals caused solely by the slight 
differences in resolution in the two epochs of data due to small, but inevitable changes in the instrumental 
set-up between epochs-exposures.

\subsection{Welty et al. data}
In some cases, the earlier epoch 1 data used for comparison were taken from \citet[][Henceforth refered to as WHK]{wel94} and \citet[][Henceforth  WMH]{wel96}. 
WHK observations of Na {\sc i} D$_1$ only,  were made in 1987-89 on a coud\'{e} spectrograph of the 2.7m telescope at McDonald observatory. 
WMH observations of Ca {\sc ii} were obtained  between 1993 and 1995 with the 0.9m coud\'{e} feed telescope and 2.1 m coud\'{e} spectrograph of the Kitt Peak National Observatory, and with the Ultra-High Resolution Facility (UHRF) on the 3.9 m Anglo-Australian Telescope in 1994. All of these epoch 1 data have very high S/N ranging 100-700 averaging around 300, and spectral resolution of $R\sim 214,000 -- 938,000$.
Twenty-one sightlines had two epochs available so the McDonald data set is therefore significantly smaller than that from UVES, but has nearly twice the spectral resolution and a significantly longer interval between epochs. 
The WHK Na{\sc i } spectra have higher resolution and so were degraded using a Gaussian of appropriate resolution, for comparison with our later observations. Data were reduced using standard methods within {\sc iraf}. Spectra were renormalized to remove stellar continua and telluric corrections were applied where necessary in the same manner as described above for the UVES data set.
\subsection{McDonald Data}
Additional observations were taken at the W.J. McDonald Observatory with the
Tull coud\'{e} echelle spectrograph on the 2.7m Harlan J. Smith telescope
\citep{tul95} during May-June 2008 and July 2014. Spectra were acquired
at two different wavelength settings. One setting provided the  principal
interstellar lines Ca\,{\sc ii} K and Ca\,{\sc i} at 4226\AA, as well
as the potential for detecting lines from K\,{\sc i} at 4044\AA\ and molecules
CH, CH$^+$ and CN. A second setting covered the interstellar Na{\sc i} D
lines. Both settings provided a resolving power $R \simeq 140,000-160,000$.\\

The final stellar sample is outlined in Table 2. The galactic coordinates, V-band magnitude, B-V colour and E(B-V) reddening (Hunter et al. 2006), proper motions, parallaxes (van Leeuwen 2007) and spectral types (taken from the POP web site (Bagnulo et al. 2003), WHK and WMH) are listed for each sightline. 
\begin{figure}
   \centering
   \includegraphics[width=\columnwidth]{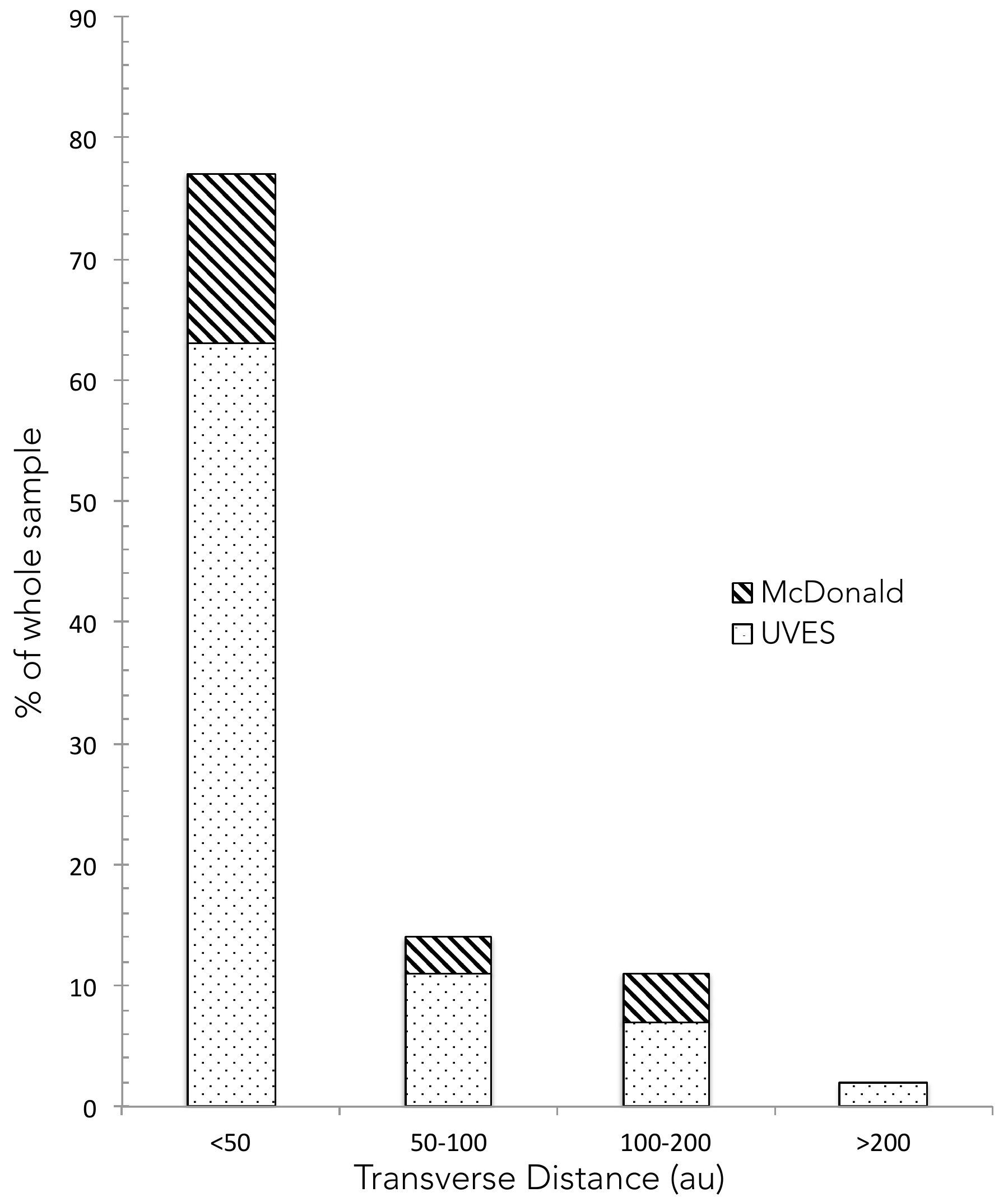}
   \caption{Histogram of transverse distance moved for the stars in the UVES and McDonald samples between the two epochs.} 
   \label{f_hist}
\end{figure}

\begin{table}
\begin{center}
\caption[]{Interstellar absorption-lines investigated in this paper, given in order of increasing wavelength. The wavelengths $\lambda_{\rm air}$ and oscillator strengths $f$ are taken from \citet{mor03} for the atomic species, \citet{bla88} for CH (mean of two unresolved transitions) and for CN, and \citet{gre93} for CH$^+$.}
\label{t_Species}
\begin{tabular}{lll}
\hline
Trans.               & $\lambda_{\rm air}$ & $f$      \\
                     &  (\AA)              &                \\
\hline
Na\,{\sc i}          & 3302.369            & 0.009        \\
Na\,{\sc i}          & 3302.978            & 0.005        \\
Ti\,{\sc ii}         & 3383.759           & 0.358          \\
Fe\,{\sc i}          & 3719.935           & 0.041         \\
Ti\,{\sc i}          & 3646.286            & 0.236          \\
Ca\,{\sc ii}  K      & 3933.661           & 0.627         \\
Ca\,{\sc ii}  H      & 3968.470            & 0.312         \\
CN                   & 3873.999            & 0.023      \\
CN                   & 3874.607            & 0.034       \\
CH$^{+}$      & 3957.692            & 0.003        \\
K\,{\sc i}           & 4044.142           & 0.006        \\
Ca\,{\sc i}          & 4226.728            & 1.770          \\
CH$^{+}$      & 4232.548            & 0.006        \\
CH                   & 4300.313            & 0.005       \\
Na\,{\sc i} D$_2$       & 5889.951          & 0.641      \\
Na\,{\sc i} D$_1$       & 5895.924           & 0.320      \\
\hline
\end{tabular}
\end{center}
\end{table}

\begin{table*}
\begin{center}
\caption{The stellar sample. Columns are as follows: the HD~number of the star; the galactic coordinates 
$l$ and $b$; the $V$ band magnitude, $B-V$ colour and reddening $E(B-V)$ \citep{hun06}; the 
proper motions in right ascension ($\mu_\alpha$) and declination ($\mu_\delta$), the parallax ($\pi$) (from \citet{van07}) 
the spectral types and the instrument used to obtain our 
data set.}
\label{t_stars}

\begin{tabular}{llllrrrrclll}

\hline
Star        & $l$	& $b$	& $V$	& $B-V$	& $E(B-V)$	& $\mu_\alpha$	& $\mu_\delta$	& $\pi$	& Spectral   & Data set    \\									
 & (deg)	& (deg)	& (mag)	& (mag)	& (mag)	& (mas\,yr$^{-1}$)	& (mas\,yr$^{-1}$)	& (mas)	& type	& \\
\hline
HD~	10840	&	291.7609	&$-$54.9427	&6.79	&	$-$0.08	&	-			&26.93	$\pm$	0.29&	$-$5.47	$\pm$	0.34		&	5.55	$\pm$	0.36			&	B9p	&	UVES\\		
HD~	29138	&	297.987	&$-$30.5423	&7.19	&	$-$0.09	&	0.1		&$-$2.53	$\pm$0.36	&8.93	$\pm$0.45				&	0.16	$\pm$	0.34		&	B1Iab	&	UVES\\					
HD~	30677	&	190.1808	&$-$22.2169	&6.84	&	$-$0.02	&	0.17		&$-$3.13	$\pm$0.57	&$-$4.00	$\pm$0.39			&	0.89	$\pm$	0.53			&	B1II/III((n)	&	UVES\\				
HD~	33328	&	209.1402	&$-$26.6856	&4.25	&	$-$0.18	&	0.01		&0.25	$\pm$0.16	&$-$1.97	$\pm$0.11				&	4.02	$\pm$	0.18		&	B2IVne	&	UVES\\					
HD~	43285	&	203.4236	&$-$05.1276	&6.05	&	$-$0.11	&	0.03		&$-$0.40	$\pm$0.42	&$-$17.42	$\pm$0.32			&	4.99	$\pm$	0.43			&	B6Ve	&	UVES\\				
HD~	45725	&	216.6614	&$-$08.2139	&4.6	&	$-$0.1	&	0.08			&$-$7.0	$\pm$1.2	&$-$5.0	$\pm$0.9			&	-				&	B3Ve	&	UVES\\						
HD~	46185	&	221.9706	&$-$10.0784	&6.79	&	$-$0.16	&	$-$0.02		&$-$4.18	$\pm$0.57	&$-$0.31	$\pm$0.48				&	2.63	$\pm$	0.55			&	B2/B3II:	&	UVES\\			
HD~	46462	&	245.4745	&$-$19.6548	&7.56	&	$-$0.10	&	0.52			&$-$8.81	$\pm$0.41	&9.43	$\pm$0.57				&	2.93	$\pm$	0.52		&	B9	&	UVES\\				
HD~	47116	&	233.174	&$-$13.9282	&7.71	&	0.01	&	0.52			&0.06	$\pm$0.44	&5.43	$\pm$0.49				&	1.53	$\pm$	0.61				&	B9p	&	UVES\\			
HD~	48099	&	206.2096	&+00.7982	&6.37	&	$-$0.08	&	0.21		&0.84	$\pm$0.41	&2.5	$\pm$0.31						&	1.17	$\pm$	0.41		&	O7V	&	UVES\\				
HD~	49131	&	240.5001	&$-$14.7267	&5.8	&	$-$0.19	&	0			&$-$2.49	$\pm$0.41	&3.80	$\pm$0.53					&	1.72	$\pm$	0.54			&	B2III	&	UVES\\				
HD~	49333	&	231.3595	&$-$10.3419	&6.08	&	$-$0.19	&	$-$0.15			&$-$14.53	$\pm$0.43	&5.12	$\pm$0.51			&	4.14	$\pm$	0.51		&	B7p	&	UVES\\				
HD~	50896	&	234.7568	&$-$10.0832	&6.74	&	$-$0.06	&	0.17		&$-$3.76	$\pm$0.33	&4.35	$\pm$0.56				&	0.72	$\pm$	0.62		&	WN5	&	UVES\\					
HD~	51876	&	228.0469	&$-$05.7456	&7.13	&	$-$0.08	&	$-$0.07			&$-$0.27	$\pm$0.48	&8.70	$\pm$0.50		&	3	$\pm$	0.59				&	B9II	&	UVES\\				
HD~	52918	&	218.0123	&+00.6139	&4.99	&	$-$0.2	&	0.03		&$-$5.05	$\pm$0.26	&2.24	$\pm$0.19					&	2.68	$\pm$	0.22		&	B1V	&	UVES\\					
HD~	58343	&	231.0872	&$-$00.2141	&5.2	&	$-$0.04	&	0.17		&$-$4.95	$\pm$0.47	&$-$20.21	$\pm$0.45				&	3.48	$\pm$	0.58		&	B2Vne	&	UVES\\				
HD~	58377	&	242.1965	&$-$06.2584	&6.81	&	$-$0.17	&	$-$0.02		&$-$7.14	$\pm$0.32	&5.69	$\pm$0.48				&	1.42	$\pm$	0.47		&	B5IV	&	UVES\\					
HD~	58978	&	237.4106	&$-$02.9978	&5.61	&	$-$0.13	&	0.06		&$-$7.76	$\pm$0.25	&4.82	$\pm$0.34				&	2.61	$\pm$	0.33		&	B1II	&	UVES	\\				
HD~	60498	&	247.1984	&$-$06.6420	&7.35	&	$-$0.12	&	0.03		&$-$8.63	$\pm$0.29	&4.18	$\pm$0.34					&	4.19	$\pm$	0.34		&	B4III	&	UVES\\				
HD~	61429	&	240.6523	&$-$01.8360	&4.7	&	$-$0.11	&	$-$0.01		&$-$3.03	$\pm$0.13	&$-$7.42	$\pm$0.22				&	5.27	$\pm$	0.26			&	B8IV	&	UVES	\\		
HD~	62714	&	268.5841	&$-$15.6252	&7.25	&	$-$0.08	&	0.05			&$-$14.30	$\pm$0.51	&7.15	$\pm$0.49				&	4.1	$\pm$	0.41		&	B7Vp	&	UVES	\\			
HD~	64972	&	245.0907	&$-$00.0164	&7.18	&	$-$0.11	&	$-$0.05		&$-$9.1	$\pm$1.5	&6.8	$\pm$1.3				&	-				&	B8p	&	UVES\\							
HD~	67536	&	276.1126	&$-$16.1399	&6.24	&	$-$0.09	&	0.12		&$-$5.36	$\pm$0.27	&11.43	$\pm$0.21				&	2.72	$\pm$0.22		&	B2.5Vn	&	UVES\\						
HD~	68761	&	254.3713	&$-$01.6188	&6.53	&	$-$0.09	&	0.13		&$-$4.44	$\pm$0.26	&5.71	$\pm$0.29				&	1.63	$\pm$	0.34			&	B0.5III	&	UVES\\				
HD~	72067	&	262.0769	&$-$03.0778	&5.83	&	$-$0.16	&	0.05		&$-$7.00	$\pm$	0.59&	6.98	$\pm$	0.41					&	2.28	$\pm$	0.49		&	B2Vne	&	UVES	\\
HD~	74966	&	258.0811	&+03.9290	&7.43	&	$-$0.14	&	0.01		&$-$10.03	$\pm$	0.36&	6.98	$\pm$	0.38			&	2.24	$\pm$	0.52		&	B4IV	&	UVES\\			
HD~	76131	&	273.5774	&$-$07.2675	&6.69	&	$-$0.05	&	0.08		&$-$16.52	$\pm$	0.30&	6.16	$\pm$	0.25		&	2.38	$\pm$	0.29			&	B6III	&	UVES	\\	
HD~	76341	&	263.5348	&+01.5210	&7.17	&	0.22	&	0.49		&$-$5.16	$\pm$	0.40&	3.49	$\pm$	0.44					&	0.75	$\pm$	0.48		&	O9.7Iab	&	UVES\\		
HD~	83312	&	284.7656	&$-$10.4703	&7.47	&	$-$0.01	&	0.15			&$-$13.47	$\pm$	0.48&	7.24	$\pm$	0.45	&	1.93	$\pm$	0.4			&	B4:psh	&	UVES\\			
HD~	88661	&	283.0814	&$-$01.4804	&5.76	&	$-$0.12	&	0.08		&$-$10.56	$\pm$	0.37&	6.81	$\pm$	0.35		&	2.56	$\pm$0.37		&	B2IVe	&	UVES\\				
HD~	89587	&	279.8352	&+05.1930	&6.87	&	$-$0.13	&	0.03		&$-$10.99	$\pm$	0.33&	6.46	$\pm$	0.31			&	0.89	$\pm$	0.38		&	B3III	&	UVES\\			
HD~	90177	&	285.1466	&$-$01.9838	&7.57	&	0.86	&	1.07			&$-$6.87	$\pm$	0.99&	2.33	$\pm$	0.77		&	1.69	$\pm$	0.82		&	B2ev	&	UVES\\				
HD~	90264	&	289.0967	&$-$08.1064	&4.95	&	$-$0.12	&	$-$0.01			&$-$22.39	$\pm$	0.20&	11.48	$\pm$	0.18	&	8.12	$\pm$	0.18			&	B8p	&	UVES\\		
HD~	90882	&	248.4149	&+44.5898	&5.18	&	$-$0.04	&	0.01		&$-$48.86	$\pm$	0.22&	$-$13.43	$\pm$	0.19			&	10.13	$\pm$	0.23		&	B9.5V	&	UVES\\			
HD~	92740	&	287.1742	&$-$00.8475	&6.42	&	0.08	&	0.08		&$-$7.71	$\pm$	0.36&	2.86	$\pm$	0.3					&	0.08	$\pm$	0.34		&	WN7A	&	UVES\\		
HD~	94910	&	289.1829	&$-$00.6953	&7.09	&	0.49	&	0.08		&$-$5.89	$\pm$	0.83&	2.45	$\pm$	0.76					&	0.19	$\pm$	0.7		&	B2:pe	&	UVES\\		
HD~	94963	&	289.7649	&$-$01.8088	&7.14	&	$-$0.1		&	0.19		&$-$7.9	$\pm$	1.3	&	2.8	$\pm$			1.3				&	-		&	O6.5IIIe	&	UVES\\		
HD~	100841	&	294.4727	&$-$01.3974	&3.12	&	$-$0.04	&	0.03		&$-$33.41	$\pm$	0.33&	$-$7.08	$\pm$	0.29		&	7.77	$\pm$0.34	&	B9III	&	UVES\\				
HD~	105056	&	298.9454	&$-$07.0551	&7.42	&	0.09	&	0.26		&$-$4.34	$\pm$	0.44&	$-$1.28	$\pm$	0.4					&	0.21	$\pm$	0.49	&	ON9.7Ia	&	UVES\\		
HD~	105071	&	298.2376	&$-$03.0911	&6.31	&	0.13	&	0.19		&$-$5.76	$\pm$	0.43&	$-$0.77	$\pm$	0.32						&	0.05	$\pm$	0.41	&	B6Ia/Ib	&	UVES\\	
HD~	106068	&	298.5069	&$-$00.4152	&5.95	&	0.23	&	0.26		&$-$5.07	$\pm$	0.30&	$-$0.52	$\pm$	0.3					&	0.08	$\pm$	0.34		&	B8Ia/Iab	&	UVES\\		
HD~	109867	&	301.7113	&$-$04.3511	&6.26	&	0		&	0.19		&$-$4.54	$\pm$	0.31&	$-$1.09	$\pm$	0.28				&	$-$0.54	$\pm$	0.32		&	B1Ia	&	UVES\\		
HD~	112272	&	303.4864	&$-$01.4947	&7.39	&	0.64	&	0.84		&$-$8.99	$\pm$	0.42&	$-$8.10	$\pm$	0.34					&	1.4	$\pm$	0.53		&	B0.5Ia	&	UVES\\		
HD~	112842	&	304.0558	&+02.4766	&7.08	&	0.15	&	0.25		&$-$4.24	$\pm$	0.45&	$-$0.95	$\pm$	0.38					&	0.31$\pm$	0.52	&	B4IV	&	UVES\\			
HD~	113904	&	304.6745	&$-$02.4907	&5.69	&	$-$0.11	&	0.11		&$-$4.26	$\pm$	0.39&	$-$2.18	$\pm$	0.48				&	0.26	$\pm$	0.48	&	WC5/B0Ia	&	UVES\\		
HD~	115363	&	305.8832	&$-$00.9683	&7.82	&	0.5	&	0.69		&$-$8.48	$\pm$	0.52&	$-$0.67	$\pm$	0.54						&	0.87	$\pm$	0.68			&	B1Ia	&	UVES\\
HD~	115842	&	307.0804	&+06.8343	&6.04	&	0.21	&	0.41		&$-$3.04	$\pm$	0.31&	3.93	$\pm$	0.36					&	0.65	$\pm$	0.44		&	B0.5Ia	&	UVES\\		
HD~	120709	&	317.2822	&+28.1893	&4.53	&	$-$0.14	&	0.01		&$-$34.54	$\pm$	0.78&	$-$28.21	$\pm$	0.62		&	9.49	$\pm$	0.89	&	B5Ip	&	UVES\\			
HD~	123515	&	315.1159	&+09.5089	&5.96	&	$-$0.04	&	0.03		&$-$32.25	$\pm$	0.32&	$-$11.63	$\pm$	0.26		&	3.98	$\pm$	0.39		&	B8p	&	UVES\\			
HD~	125823	&	321.5656	&+20.0226	&4.40	&	$-$0.17	&	$-$0.05		&$-$24.15	$\pm$	0.13&	$-$21.90	$\pm$	0.13		&	7.13	$\pm$	0.16		&	B8p	&	UVES\\			
HD~	133518	&	323.0822	&+05.4606	&6.40	&	$-$0.11	&	0.08		&$-$2.83	$\pm$	0.40&	$-$12.52	$\pm$	0.48			&	2.21	$\pm$	0.44		&	B2IV 	&	UVES\\		
HD~	136239	&	321.2281	&$-$01.7448	&7.87	&	0.77	&	0.94		&$-$5.70	$\pm$	0.99&	$-$2.20	$\pm$	0.9					&	0.09	$\pm$	0.9		&	B5p	&	UVES\\		
HD~	137509	&	315.2529	&$-$12.1253	&6.89	&	$-$0.13	&	$-$0.02		&$-$16.08	$\pm$	0.28&	$-$14.64	$\pm$	0.4		&	5.11	$\pm$	0.38		&	B8p	&	UVES\\			
HD~	137753	&	325.9312	&+03.3436	&6.7	&	$-$0.03	&	0.09		&$-$7.79	$\pm$	0.39&	$-$8.97	$\pm$	0.38					&	3.3	$\pm$	0.42		&	B7IV	&	UVES\\		
HD~	142301	&	347.1237	&+21.5122	&5.86	&	$-$0.07	&	0.04		&$-$13.16	$\pm$	0.60&	$-$25.55	$\pm$	0.46				&	6.33	$\pm$	0.43		&	B8p	&	UVES\\		
\hline
\end{tabular}
\end{center}
\end{table*}

\addtocounter{table}{-1}
\begin{table*}
\caption{continued}
\label{t_stars_ctd}
\begin{tabular}{llllrrrrclll}
\hline
Star        & $l$	& $b$	& $V$	& $B-V$	& $$E(B-V)$$	& $\mu_\alpha$	& $\mu_\delta$	& $\pi$	& Spectral   & Data set    \\	
								 & (deg)	& (deg)	& (mag)	& (mag)	& (mag)	& (mas\,yr$^{-1}$)	& (mas\,yr$^{-1}$)	& (mas)		&type & \\
\hline
HD~	142758	&	325.3065	&$-$04.2765	&7.08	&	0.14	&	0.31		&$-$5.38	$\pm$	0.48&	$-$5.33	$\pm$	0.47					&	$-$0.76	$\pm$	0.51	&	B1.5Ia	&	UVES\\		
HD~	142983	&	356.3867	&+28.6317	&2.29	&	$-$0.09	&	0.14		&$-$12.44	$\pm$	0.25&	$-$16.73	$\pm$	0.21		&	6.97	$\pm$	0.24	&	B8Ia/Iab	&	UVES\\			
HD~	143448	&	324.5485	&$-$05.9714	&7.30	&	$-$0.00	&	0.16		&$-$3.88	$\pm$	0.54&	$-$2.72	$\pm$	0.55				&	1.54	$\pm$	0.58		&	B3IVe	&	UVES\\		
HD~	145482	&	348.1166	&+16.8354	&4.57	&	$-$0.15	&	0.06		&$-$10.38	$\pm$	0.18&	$-$23.94	$\pm$	0.14		&	6.81	$\pm$	0.16			&	B2V	&	UVES\\		
HD~	145792	&	351.0096	&+19.0290	&6.42	&	0.08	&	0.14		&$-$8.89	$\pm$	0.71&	$-$20.23	$\pm$	0.47			&	6.96	$\pm$	0.65		&	B6p	&	UVES\\			
HD~	145842	&	334.7002	&+02.5325	&5.12	&	$-$0.10	&	0.01		&$-$34.39	$\pm$	0.25&	$-$44.88	$\pm$	0.22			&	8.46	$\pm$	0.26		&	B8V	&	UVES\\		
HD~	148184	&	357.9328	&+20.6766	&4.28	&	0.31	&	0.52		&$-$5.41	$\pm$	0.31&	$-$21.12	$\pm$	0.24					&	6.21	$\pm$	0.23	&	B2Vne	&	UVES\\		
HD~	148379	&	337.2456	&+01.5757	&5.36	&	0.44	&	0.61		&$-$4.54	$\pm$	1.06&	$-$1.51	$\pm$	1.27					&	1.79	$\pm$	0.7	&	B1.5Iae	&	UVES\\		
HD~	148688	&	340.7197	&+04.3474	&5.33	&	0.27	&	0.46		&$-$4.35	$\pm$	0.37&	$-$1.37	$\pm$	0.29					&	1.2	$\pm$	0.33	&	B1Ia	&	UVES\\		
HD~	148937	&	336.3661	&$-$00.2181	&6.77	&	0.24	&	0.09		&1.51	$\pm$	0.86&	$-$3.57	$\pm$	0.77					&	2.35	$\pm$	0.79	&	O6.5e	&	UVES\\		
HD~	152003	&	343.3324	&+01.4077	&6.99	&	0.29	&	0.53		&$-$2.5	$\pm$	1	&	$-$2.5	$\pm$	1						&	-				&	O9Iab	&	UVES\\	
HD~	152235	&	343.3111	&+01.1041	&6.34	&	0.42	&	0.61		&$-$1.07	$\pm$	0.56&	$-$3.30	$\pm$	0.39					&	1.05	$\pm$	0.57	&	B0.7Ia	&	UVES\\		
HD~	154811	&	341.0616	&$-$04.2191	&6.93	&	0.4	&	0.46		&$-$1.13	$\pm$	0.62&	$-$3.86	$\pm$	0.43						&	2.34	$\pm$	0.69	&	OC9.7Iab	&	UVES\\	
HD~	154873	&	341.3454	&$-$04.1096	&6.70	&	0.28	&	0.47		&1.47	$\pm$	0.72&	$-$4.83	$\pm$	0.57					&	2.16	$\pm$	0.72		&	B1Ib	&	UVES\\		
HD~	155416	&	348.7532	&+00.6515	&6.90	&	0.3	&	0.34		&$-$0.26	$\pm$	0.59&	$-$2.32	$\pm$	0.27					&	0.9	$\pm$	0.54	&	B7II	&	UVES\\		
HD~	155806	&	352.5858	&+02.8683	&5.61	&	$-$0.06	&	0.23		&0.92	$\pm$	0.36&	$-$3.05	$\pm$	0.19				&	0.89	$\pm$	0.4		&	O7.5Ve	&	UVES\\		
HD~	156385	&	343.1641	&$-$04.7627	&6.92	&	0.05	&	0.35		&1.75	$\pm$	0.61&	$-$3.89	$\pm$	0.3					&	$-$0.26	$\pm$	0.57		&	WC7	&	UVES\\		
HD~	157038	&	349.9548	&$-$00.7940	&6.41	&	0.64	&	0.81		&$-$1.33	$\pm$	0.52&	$-$0.07	$\pm$	0.25					&	0.51	$\pm$	0.42		&	B1/B2Ia	&	UVES\\		
HD~	163745	&	350.5576	&$-$08.7882	&6.5	&	$-$0.09	&	$-$0.01		&1.43	$\pm$	0.52&	$-$3.77	$\pm$	0.21				&	2.63	$\pm$	0.4		&	B5II	&	UVES\\		
HD~	163758	&	355.3605	&$-$06.1042	&7.32	&	0.02	&	0.34		&3.3	$\pm$	1.3	&	$-$4.8	$\pm$	1.2					&	-				&	O6.5Iaf	&	UVES\\		
HD~	163800	&	7.052	&+00.6878	&7.01	&	0.31	&	0.6		&$-$0.01	$\pm$	0.84&	$-$1.80	$\pm$	0.36						&	$-$1.01	$\pm$0.66			&	O7III	&	UVES\\	
HD~	167264	&	10.4557	&$-$01.7408	&5.35	&	0		&	0.2		&1.60	$\pm$	0.57&	$-$1.51	$\pm$	0.41				&	0.1	$\pm$	0.45		&	B0.5Ia/Iab	&	UVES\\		
HD~	169454	&	17.5385	&$-$00.6697	&6.65	&	0.74	&	0.93		&2.21	$\pm$	0.87&	$-$4.49	$\pm$	0.61					&	$-$1.13	$\pm$	0.72	&	B1	Ia	Ce	&	UVES\\
HD~	170235	&	7.9523	&$-$06.7282	&6.59	&	0.07	&	0.28		&$-$1.00	$\pm$	0.62&	$-$6.37	$\pm$	0.39					&	1.07	$\pm$	0.61		&	B2Vnne	&	UVES\\		
HD~	188294	&	32.6528	&$-$17.7667	&6.44	&	$-$0.02	&	0.09		&2.13	$\pm$	0.49&	$-$30.77	$\pm$	0.46				&	6.42	$\pm$	0.58		&	B8V	&	UVES\\		
HD~	199728	&	28.537	&$-$36.5172	&6.25	&	$-$0.13	&	-		&10.07	$\pm$	0.44&	$-$20.09	$\pm$	0.36					&	6.09	$\pm$	0.42		&	B9p	&	UVES\\	
HD~	221507	&	355.0398	&$-$70.3628	&4.37	&	$-$0.08	&	-		&95.97	$\pm$	0.10&	38.29	$\pm$	0.12					&	18.74	$\pm$	0.15		&	B9.5IVmnp	...&	UVES\\		
HD~	223640	&	60.5398	&$-$73.9377	&5.18	&	$-$0.12	&	-		&26.82	$\pm$	0.38&	$-$4.27	$\pm$	0.3			&	10.23	$\pm$	0.31		&	Ap...	&	UVES\\			
HD~	91316		&	234.8870	&+52.7674	&	3.870			&$-$0.15	&0.04				&	$-$5.93		$\pm$0.20	&$-$3.40	$\pm$0.11	&	0.60	$\pm$	0.18	&	B1Iab	&	McD \\				
	
HD~	114330		&	311.4176	&+57.0274	&	4.381			&0		&$-$0.01				&	-36.28		$\pm$0.93	&-31.22	$\pm$0.69	&	10.33	$\pm$	1.09		&	A1IVs &	 McD 	\\		

HD~	141637		&	346.0979	&+21.7059	&	4.638			&$-$0.072	&0.15				&	$-$14.20		$\pm$0.33	&$-$25.12	$\pm$0.26	&	6.59	$\pm$	0.27		&	B1.5Vn	&	McD 	\\	
HD~	143018		&	347.2148	&+20.2305	&	2.89				&$-$0.19	&0.04				&	$-$11.42		$\pm$0.78	&$-$26.83	$\pm$0.74	&	5.57	$\pm$	0.64		&	B1V	&	McD \\	
HD~	143275		&	350.0969	&+22.4904	&	2.291			&-0.086	&0.14				&	-10.21		$\pm$1.01	&-35.41	$\pm$0.71	&	6.64	$\pm$	0.89	&		B0.2IVe	&	McD	\\		
HD~	144217		&	353.1929	&+23.5996	&	2.62				&$-$0.07	&0.17				&	$-$5.20		$\pm$0.92	&$-$24.04	$\pm$0.64	&	8.07	$\pm$	0.78		&	B0.5V	&	McD	\\
HD~	144470		&	352.7497	&+22.7730	&	3.946			&-0.046	&0.18				&	-8.98		$\pm$0.23	&-23.48	$\pm$0.16	&	6.92	$\pm$	0.26		&	B1V	&	McD 	\\	
HD~	147084		&	352.3279	&+18.0503	&	4.55				&0.84	&0.24				&	-4.32		$\pm$0.52	&-14.15	$\pm$0.40	&	3.71	$\pm$	0.54		&	A4II/III	&	McD \\	
HD~	147165		&	351.3130	&+16.9989	&	2.912			&0.097	&0.31				&	-10.6		$\pm$0.78	&-16.28	$\pm$0.43	&	4.68	$\pm$	0.60		&	B2III	&	McD 	\\	
HD~	148184		&	357.9328	&+20.6766	&	4.42				&0.28	&0.49				&	$-$5.41		$\pm$0.31	&$-$21.12	$\pm$0.24		&	6.21	$\pm$	0.23	&		B1.5Ve	&	McD 	\\	
HD~	149438		&	351.5348	&+12.8081	&	2.814			&$-$0.201	&0.49				&	$-$9.89		$\pm$0.61	&$-$22.83	$\pm$0.55		&	6.88	$\pm$	0.53		&	B0V	&	McD	\\		
HD~	149757		&	006.2812	&+23.5877	&	2.578			&0.017	&0.30				&	15.26		$\pm$0.26	&24.79		$\pm$0.22		&	8.91	$\pm$	0.20		&	O9.5Vnn	&	McD 	\\	
HD~	152614		&	028.7341	&+30.6605	&	4.38				&-0.08	&0.03				&	-53.8		$\pm$0.22	&-34.04	$\pm$0.19	&	13.3	$\pm$	0.22		&	B8V	&	McD \\	

HD~	159561		&	035.8937	&+22.5681	&	2.1				&0.13	&0.03				&	108.07		$\pm$0.87	&$-$221.57$\pm$0.80	&	67.13	$\pm$	1.06		&	A5III	&	McD 	\\	
HD~	184915		&	031.7709	&-13.2866	&	4.960			&-0.03	&0.19				&	1.63			$\pm$0.20	&-2.65		$\pm$0.10	&	1.94	$\pm$	0.20		&	B0.5III	&	McD 	\\	
HD~	186882		&	078.7096	&+10.2430	&	2.900			&$-$0.019	&0.02				&	44.07		$\pm$0.46	&48.66		$\pm$0.49			&	19.77	$\pm$	0.48	&	B9.5III	&	McD 	\\	
HD~	197345		&	084.2847	&+01.9975	&	1.25				&0.09	&0.02				&	2.01			$\pm$0.34	&1.85		$\pm$0.27	&	2.31	$\pm$	0.32	&	A2Iae	&	McD 	\\
HD~	198183		&	078.0844	&$-$04.3378	&	4.75				&$-$0.12	&0.03				&	14.71		$\pm$0.32	&$-$8.96			$\pm$0.49	&	4.24	$\pm$	0.43		&	B6IV	&	McD \\	
HD~	200120		&	088.0296	&+00.9707	&	4.74				&$-$0.05	&0.17				&	8.12			$\pm$0.42	&2.15		$\pm$0.50	&	2.3	$\pm$	0.42	&		B1.5Vnne	&	McD	\\	
HD~	202904		&	080.9767	&$-$10.0526	&	4.43				&$-$0.11	&0.10				&	10.03		$\pm$0.33	&6.49			$\pm$0.38	&	5.08	$\pm$	0.55	&		B2Ve	&	McD\\		
HD~	212571		&	066.0067	&$-$44.7394	&	4.794			&$-$0.168	&0.06				&	17.83		$\pm$0.27	&2.41			$\pm$0.20	&	4.17	$\pm$	0.28	&	B1Ve	&	McD\\			
                                                                                                                                                                            
\hline
\end{tabular}

\end{table*}

\subsection{Stellar distances and proper motions}
\label{s_distance}
Stellar distances, $(B-V)$ and $E(B-V)$ values were, in most cases, taken from \citet{van07} and \citet{hun06}, taking the most accurate measurement in each case. For some sightlines neither \citet{hun06} nor \citet{van07} gave distance estimates and so values were taken from the literature. These cases are referenced in Table \ref{t_dist}. Most distances were calculated from the parallax recorded in \cite{van07} as these tended to be the most reliable. The distance estimates from \citet{hun06} were calculated using the spectroscopic parallax methodology given in \cite{dip94}.  The distances, along with the proper motions of the stars \citep{van07} were used to calculate the transverse displacement on the sky during the interval between epochs by trigonometry, to allow us to set some limits on the scale over which any variation in the ISM occurred. We must note that there is no way to account for the motion of the clouds themselves which could act to increase or decrease this estimate on the scale of the structure in question, for example, if we assume the cloud itself to  have a transverse velocity of 5\kms, then in 6 yr it will have moved $\approx$ 6 au, and depending on the direction of this motion, it could increase or decrease our transverse distance estimates calculated using the proper motion of the background star. We can therefore only say that the distance the star has moved across the sky is an indicator of the true scale of the ISM structure. Table~\ref{t_stars} lists our final sample, whilst Fig.~\ref{f_hist} shows a histogram of the transverse distance moved for the 104 stars between the two epochs. 

\subsection{Column Densities}
\label{s_columndensity}

For sightlines where variations were detected, column densities were
calculated using the {\sc vapid} suite \citep{how02}. This assumes Gaussian
line of sight velocity distributions of pure absorbers in each cloud
\citep{str48}, and so can obtain values for column density ($N$),
central velocity ($v$) and velocity dispersion ($b$) for each
component of a specified absorption profile by fitting multiple Voigt profiles.
Even at the high spectral resolution of the current study it is likely 
that most components will be unresolved at least in the UVES data set. 
WHK investigated the statistics of line widths and component separations at a significantly higher resolution than our UVES data. They found a median $b$(Na {\sc i}) of 0.73 \kms and median separation between adjacent components of about 2 \kms (their figs. 5 and 7). Even at such high resolutions, these values are expected to be over estimates of the true values, and so the likelihood of unresolved structure in our UVES data is very high. 

For Na I, the D$_1$ and D$_2$ lines were fitted simultaneously, as were the H and K lines of Ca II. 
As very few velocity components showed variation between epochs, 
these stable components were measured in epoch one and fixed when fitting epoch
2, allowing us to accurately quantify the change in column densities in the varying
components.

Column densities for all varying sightlines were recalculated, changing the instrumental resolution by 10 \% and fixing the velocity values within {\sc vapid}, allowing the $b$-value and column density to vary freely. In most cases the differences derived due to the change in resolution were negligible, and well within the errors presented for the column densities. The only sightlines where larger variations due to this change in resolution were found, were in the particularly narrow components. However, even in these cases the variation caused by changing the instrumental $b$-value was small when compared with the variation seen between epochs.

\section{Results}        
\label{s_results}        

In this section the results of the search for time variability in the twin-epoch spectra are discussed. 
We note that O- and B-type stars have high multiplicity \citep{mas09,san13}, and so great care must be taken not to confuse variations in stellar or circumstellar features with interstellar ones.
Each pair of spectra and their residuals in the transitions listed in Table~\ref{t_Species} were searched by eye for evidence of differences between epochs. Only the strong lines of Na\,{\sc i}~D, Ca\,{\sc ii}~H\&K and the weaker Ca\,{\sc i} showed convincing evidence for significant interstellar variation. The Na\,{\sc i} D lines and Ca\,{\sc ii} H\&K transitions each form a doublet, so putative variations were required to be present in both components of the doublet before they were accepted as real. For Ca\,{\sc i}, the variations are always seen at the same velocity as corresponding variations in Na\,{\sc i} or Ca\,{\sc ii} which provides confidence in their reality.

Of the total 83 twin-epoch UVES spectra analysed, 80\% show no significant difference between the two epochs in any species. 2\% show variation caused by eclipsing 
companions, 12\%  show changes that are probably due to circumstellar material, 
and 6\% show possible time variation in one or more velocity component caused by interstellar absorption - typically there are $\ga$ 5 velocity components per sightline. The sightlines where stellar or circumstellar variation are visible could also exhibit interstellar variations but they would be hidden due to the more prominent stellar-circumstellar profile changes. 
For the McDonald data, none of the sightlines show any substantial or convincing interstellar time variation.  The following subsections 
discuss each of the time-varying spectra.

\begin{table*}
\begin{center}
\caption{\textbf{Observation} dates, derived distances $d$ and transverse displacements for all stars in our sample. For some objects distances have been taken from the literature. References for these are given below the table.}
\label{t_dist}
\begin{tabular}{llcccrr}
\hline
Star & \multicolumn{2}{c} {Distance ($d$)}  &Epoch 1 & epoch 2 & Interval  & Traversed distance\\ 
 &{(pc)}  & ref. &date of observation  &date of observation & (yr)  & (au)\\
 
\hline 

HD~10840	&	180	$\pm$ 	11.7	& p	&	14-09-2002	 &	03-06-2008	&	5.7		&	28.3	$\pm$	1.9	\\
HD~29138	&	3530	$\pm$ 	1059	& s	&	21-01-2002	&	20-06-2008	&	6.4		&	210	$\pm$	63.9	\\
HD~30677	&	1124	$\pm$	669	& p	&	28-02-2002	&	26-07-2008	&	6.4		&	36.6	$\pm$	22.0	\\
HD~33328	&	249	$\pm$	11	& p	&	15-10-2002	&	04-04-2008	&	5.5		&	2.70	$\pm$	0.2	\\
HD~43285	&	200	$\pm$	17	& p	&	23-02-2002	&	28-08-2008	&	6.5		&	22.8	$\pm$	2.0	\\
HD~45725	&	184	$\pm$ 	55	& s	&	13-03-2001	&	04-04-2008	&	7.1		&	11.2	$\pm$	3.6	\\
HD~46185	&	380	$\pm$	80	& p	&	16-02-2002	&	14-08-2008	&	6.5		&	10.4	$\pm$	2.6	\\
HD~46462	&	341	$\pm$	61	& p	&	25-02-2002	&	14-08-2008	&	6.5		&	28.5	$\pm$	5.2	\\
HD~47116	&	654	$\pm$	261	& p	&	25-02-2002	&	02-09-2008	&	6.5		&	23.2	$\pm$	9.5	\\
HD~48099	&	855	$\pm$	300	& p	&	09-04-2001	&	19-09-2008	&	7.5		&	16.8	$\pm$	6.2	\\
HD~49131	&	581	$\pm$	183	& p	&	02-10-2002	&	18-08-2008	&	5.9		&	15.5	$\pm$	5.2	\\
HD~49333	&	242	$\pm$	30	& p	&	13-09-2002	&	18-08-2008	&	5.9		&	22.1	$\pm$	2.8	\\
HD~50896	&	1389 $\pm$	1196	& p	&	20-01-2002	&	17-09-2008	&	6.7		&	53.2	$\pm$	46.0	\\
HD~51876	&	333	$\pm$	66	& p	&	14-02-2003	&	13-09-2008	&	5.6		&	16.2	$\pm$	3.3	\\
HD~52918	&	373	$\pm$	31	& p	&	26-10-2002	&	04-04-2008	&	5.4		&	11.2	$\pm$	1.1	\\
HD~58343	&	287	$\pm$	48	& p	&	13-03-2001	&	04-04-2008	&	7.1		&	42.2	$\pm$	7.1	\\
HD~58377	&	448	$\pm$ 	134	& s	&	14-02-2003	&	03-09-2008	&	5.6		&	22.7	$\pm$	6.9	\\
HD~58978	&	383	$\pm$	48	& p	&	03-10-2002	&	21-09-2008	&	6.0		&	20.9	$\pm$	2.7	\\
HD~60498	&	239	$\pm$	19	& p	&	15-02-2003	&	19-09-2008	&	5.6		&	12.8	$\pm$	1.1	\\
HD~61429	&	190	$\pm$	9.4	& p	&	29-11-2001	&	04-04-2008	&	6.4		&	9.66	$\pm$	0.5	\\
HD~62714	&	244	$\pm$	24	& p	&	25-02-2002	&	17-09-2008	&	6.6		&	25.6	$\pm$	2.7	\\
HD~64972	&	2858 $\pm$ 	857 & s		&	27-11-2002	&	17-04-2008	&	5.4		&	175	$\pm$	56.9	\\
HD~67536	&	368	$\pm$	30	& p	&	26-10-2002	&	17-04-2008	&	5.5		&	25.4	$\pm$	2.1	\\
HD~68761	&	613	$\pm$	128	& p	&	25-02-2002	&	17-04-2008	&	6.1		&	27.3	$\pm$	5.8	\\
HD~72067	&	439	$\pm$	94	& p	&	17-11-2001	&	17-04-2008	&	6.4		&	27.8	$\pm$	6.2	\\
HD~74966	&	446	$\pm$	104	& p	&	11-12-2002	&	17-04-2008	&	5.4		&	29.2	$\pm$	6.8	\\
HD~76131	&	420	$\pm$	51	& p	&	15-02-2003	&	17-04-2008	&	5.2		&	38.3	$\pm$	4.7	\\
HD~76341	&	2365	$\pm$ 	709	& s	&	24-02-2002	&	17-04-2008	&	6.1		&	90.6	$\pm$	27.8	\\
HD~83312	&	518	$\pm$	107	& p	&	15-01-2003	&	17-04-2008	&	5.3		&	41.7	$\pm$	8.7	\\
HD~88661	&	391	$\pm$	56	& p	&	08-02-2001	&	17-04-2008	&	7.2		&	35.3	$\pm$	5.2	\\
HD~89587	&	900	$\pm$ 	270	& s	&	30-11-2001	&	17-04-2008	&	6.4		&	73.2	$\pm$	22.1	\\
HD~90177	&	255	$\pm$ 	76	& s	&	08-10-2002	&	07-06-2008	&	5.7		&	10.5	$\pm$	3.4	\\
HD~90264	&	123	$\pm$	2.7	& p	&	26-02-2002	&	03-04-2008	&	6.1		&	18.9	$\pm$	0.4	\\
HD~90882	&	99	$\pm$	2.2	& p	&	24-11-2001	&	03-04-2008	&	6.4		&	31.8	$\pm$	0.7	\\
HD~92740	&	2600	$\pm$ 	--   & b		&	02-01-2003	&	17-04-2008	&	5.3		&	113	$\pm$	--	\\
HD~94910	&	6000	$\pm$ 	1000  & c		&	11-01-2003	&	08-06-2008	&	5.4		&	207	$\pm$	43.6	\\
HD~94963	&	3257	$\pm$ 	977	& s	&	25-02-2002	&	17-04-2008	&	6.1		&	168	$\pm$	56.7	\\
HD~100841	&	129	$\pm$	5.6	& p	&	23-06-2001	&	17-04-2008	&	6.8		&	30.0	$\pm$	1.3	\\
HD~105056	&	5184	$\pm$ 	-- & b		&	23-02-2002	&	26-06-2008	&	6.3		&	149	$\pm$	--\\
HD~105071	&	3482	$\pm$ 	1044 & s		&	26-02-2002	&	17-04-2008	&	6.1		&	124	$\pm$	38.4	\\
HD~106068	&	2824	$\pm$ 	847	& s	&	20-07-2001	&	17-04-2008	&	6.7		&	97.1	$\pm$	29.7	\\
HD~109867	&	3264	$\pm$ 	979	& s	&	26-02-2002	&	17-04-2008	&	6.1		&	93.6	$\pm$	28.7	\\
HD~112272	&	714	$\pm$	270	& p	&	26-02-2002	&	17-04-2008	&	6.1		&	53.1	$\pm$	20.2	\\
HD~112842	&	2126	$\pm$ 	638	& s	&	15-02-2003	&	17-04-2008	&	5.2		&	47.8	$\pm$	15.2	\\
HD~113904	&	2817	$\pm$ 	845	& s	&	22-07-2001	&	17-04-2008	&	6.7		&	90.9	$\pm$	28.4	\\
HD~115363	&	3282	$\pm$ 	984	& s	&	26-02-2002	&	26-06-2008	&	6.3		&	177	$\pm$	54.1	\\
HD~115842	&	2157	$\pm$ 	647	& s	&	25-02-2002	&	17-04-2008	&	6.1		&	65.9	$\pm$	20.3	\\
HD~120709	&	105	$\pm$	10	& p	&	24-06-2001	&	03-04-2008	&	6.8		&	31.9	$\pm$	3.0	\\
HD~123515	&	251	$\pm$	25	& p	&	05-08-2002	&	17-04-2008	&	5.7		&	49.1	$\pm$	4.8	\\
HD~125823	&	140	$\pm$	3.1	& p	&	13-03-2001	&	03-04-2008	&	7.1		&	32.3	$\pm$	0.7	\\
HD~133518	&	452	$\pm$	90	& p	&	13-03-2001	&	17-04-2008	&	7.1		&	41.2	$\pm$	8.4	\\
HD~136239	&	2350	$\pm$ 	705	& s	&	22-02-2002	&	17-04-2008	&	6.2		&	88.4	$\pm$	30.0	\\
HD~137509	&	196	$\pm$	15	& p	&	11-07-2001	&	17-04-2008	&	6.8		&	28.8	$\pm$	2.2	\\
HD~137753	&	303	$\pm$	39	& p	&	14-02-2003	&	17-04-2008	&	5.2		&	18.6	$\pm$	2.4	\\
HD~142301	&	158	$\pm$	11	& p	&	13-03-2001	&	09-04-2008	&	7.1		&	32.1	$\pm$	2.3	\\
HD~142758	&	1316	$\pm$	883	& p	&	26-02-2002	&	17-04-2008	&	6.1		&	61.2	$\pm$	41.3	\\
\hline
\end{tabular}

Distance reference codes: p - distances calculated from the parallax recorded in \cite{van07}.\\
s - distances calculated spectroscopically \citep{hun06}. \\

b - \citet{dip94},
c - \citet{hoe93},
d - \citet{con90}
\end{center}
\end{table*}

\addtocounter{table}{-1}
\begin{table*}
\begin{center}
\caption{continued}
\label{t_dist_ctd}
\begin{tabular}{llcccrr}
\hline
Star & \multicolumn{2}{c} {Distance ($d$)}  &Epoch 1 & epoch 2 & Interval  & Traversed distance\\ 
 &{(pc)}  & ref. &date of observation  &date of observation & (yr)  & (au)\\
\hline
HD~142983	&	143	$\pm$	4.9	& p	&	11-07-2001	&	03-04-2008	&	6.7		&	20.1	$\pm$	0.7	\\
HD~143448	&	649	$\pm$	245	& p	&	26-02-2002	&	17-04-2008	&	6.1		&	18.9	$\pm$	7.4	\\
HD~145482	&	147	$\pm$	3.5	& p	&	13-08-2002	&	09-04-2008	&	5.7		&	21.7	$\pm$	0.5	\\
HD~145792	&	144	$\pm$	13	& p	&	06-10-2002	&	09-04-2008	&	5.5		&	17.5	$\pm$	1.7	\\
HD~145842	&	118	$\pm$	3.6	& p	&	26-02-2002	&	03-04-2008	&	6.1		&	40.8	$\pm$	1.3	\\
HD~148184	&	161	$\pm$	6.0	& p	&	26-02-2002	&	03-04-2008	&	6.1		&	21.4	$\pm$	0.8	\\
HD~148379	&	559	$\pm$	218	& p	&	26-02-2002	&	09-04-2008	&	6.1		&	16.4	$\pm$	7.4	\\
HD~148688	&	833	$\pm$	229	& p	&	14-09-2002	&	09-04-2008	&	5.6		&	21.2	$\pm$	6.1	\\
HD~148937	&	426	$\pm$	143	& p	&	26-02-2002	&	09-04-2008	&	6.1		&	10.1	$\pm$	4.0	\\
HD~152003	&	2687	$\pm$ 	806	& s	&	26-02-2002	&	09-04-2008	&	6.1		&	58.1	$\pm$	24.0	\\
HD~152235	&	952	$\pm$	517	& p	&	03-10-2002	&	09-04-2008	&	5.5		&	18.2	$\pm$	10.1	\\
HD~154811	&	427	$\pm$	126	& p	&	03-09-2002	&	09-04-2008	&	5.6		&	9.63	$\pm$	3.0	\\
HD~154873	&	463	$\pm$	154	& p	&	07-10-2002	&	09-04-2008	&	5.5		&	12.9	$\pm$	4.5	\\
HD~155416	&	1278	$\pm$ 	383	& s	&	27-07-2001	&	09-04-2008	&	6.7		&	20.0	$\pm$	6.4	\\
HD~155806	&	1064	$\pm$ 	319	& s	&	11-07-2001	&	09-04-2008	&	6.8		&	22.9	$\pm$	7.0	\\
HD~156385	&	1300	$\pm$ 	-- &d		&	14-10-2002	&	17-04-2008	&	5.5		&	30.6	$\pm$	--	\\
HD~157038	&	1444	$\pm$ 	433	& s	&	22-09-2002	&	03-06-2008	&	5.7		&	11.0	$\pm$	5.4	\\
HD~163745	&	380	$\pm$	58	& p	&	19-06-2001	&	27-05-2008	&	6.9		&	10.6	$\pm$	1.8	\\
HD~163758	&	4103	$\pm$ 	1231& s		&	27-07-2001	&	09-04-2008	&	6.7		&	160	$\pm$	58.9	\\
HD~163800	&	1621	$\pm$ 	486	& s	&	29-09-2002	&	23-06-2008	&	5.7		&	16.7	$\pm$	6.0	\\
HD~167264	&	2109	$\pm$ 	632	& s	&	12-07-2001	&	27-05-2008	&	6.9		&	31.9	$\pm$	12.0	\\
HD~169454	&	1359	$\pm$ 	407	& s	&	13-09-2002	&	23-06-2008	&	5.8		&	39.3	$\pm$	12.9	\\
HD~170235	&	501	$\pm$ 	150		& s&	10-07-2001	&	19-04-2008	&	6.8		&	21.9	$\pm$	7.0	\\
HD~188294	&	156	$\pm$	14	& p	&	12-07-2001	&	23-04-2008	&	6.8		&	32.6	$\pm$	3.0	\\
HD~199728	&	164	$\pm$ 	11	& p	&	29-11-2002	&	23-06-2008	&	5.6		&	20.6	$\pm$	1.5	\\
HD~221507	&	53.4	$\pm$ 	0.4	& p	&	09-06-2002	&	27-05-2008	&	6.0		&	32.9	$\pm$	0.3	\\
HD~223640	&	97.8	$\pm$ 	3	& p	&	13-08-2002	&	03-06-2008	&	5.8		&	15.4	$\pm$	0.5	\\
HD~091316   &		1346	$\pm$ 	403	& s	&	--	&	29-05-2008	&	20.0	$^*$	&	184	$\pm$	55.3	\\
HD~114330	&	97	$\pm$	10	& p 	&	27-05-2008	& 	04-07-2014	&	6.1		&	28.3	$\pm$	3.0	\\
HD~141637   &		152	$\pm$	6.2	& p 	&	23-06-1988	&	28-05-2008	&	19.9		&	167	$\pm$	50.1	\\
HD~143018   &		180	$\pm$	21	& p 	&	21-06-1988	&	28-05-2008	&	19.9		&	104	$\pm$	30.9	\\
HD~143275   &		151	$\pm$	20	& p 	&	31-05-2008	 &	06-07-2014	 &	6.1		 &	38.9	$\pm$	11.7	\\
HD~144217   &		124	$\pm$	12	& p 	&	--	&	27-05-2008	&	20.0	$^*$	&	77.2	$\pm$	23.2	\\
HD~144470   &		145	$\pm$	5.4	& p 	&	01-05-2008	&	06-07-2014	&	6.2		&	37.2	$\pm$	11.0	\\

HD~147084   &		270	$\pm$	39	& p 	&	28-05-2008	& 	04-07-2014	&	6.1		&	24.3	$\pm$	3.6	\\
HD~147165   &		214	$\pm$	27	& p 	&	28-05-2008	& 	04-07-2014	&	6.1		&	22.1	$\pm$	6.6	\\
HD~148184   &		161	$\pm$	6.0	& p 	&	30-05-2008	 &	04-07-2014	&	6.1		&	21.4	$\pm$	0.8	\\
HD~149438   &		145	$\pm$	11	& p 	&	21-06-1988	&	30-05-2008	&	20.0		&	72.0	$\pm$	5.7	\\
HD~149757   &		112	$\pm$	2.5	& p 	&	10-06-1905	&	28-05-2008	&	20.0	 $^*$	&	116	$\pm$	34.4	\\
HD~152614   &		75	$\pm$	1.2	& p 	&	30-05-2008	 &	06-07-2014	 &	6.1		&	29.2	$\pm$	0.5	\\
HD~159561   &		15	$\pm$	0.2	& p 	&	28-05-2008	&	06-07-2014	&	6.1		&	22.6	$\pm$	22.6	\\
HD~184915   &		515	$\pm$	53	& p	&	27-05-2008	& 	04-07-2014	 & 	6.1		&	12.4	$\pm$	3.7	\\
HD~186882   &		51	$\pm$	1.2	& p	&	--	&	27-05-2008	&	20.0	$^*$	&	49.9	$\pm$	14.4	\\
HD~197345   &		433	$\pm$	60	& p	&	29-05-2008	&	06-07-2014	&	6.1		&	7.22	$\pm$	1.29	\\
HD~198183   &		236	$\pm$	24	& p	&	29-05-2008	&	06-07-2014	&	6.1		&	24.8	$\pm$	2.6	\\
HD~200120   &		435	$\pm$	79	& p	&	23-06-1988	&	31-05-2008	&	20.0		&	49.6	$\pm$	15.1	\\
HD~202904   &		197	$\pm$	21	& p	&	11-10-1989	&	30-05-2008	&	18.6		&	53.2	$\pm$	15.9	\\
HD~212571   &		240	$\pm$	16	& p	&	31-05-2008	&	06-07-2014	&	6.1		&	45.8	$\pm$	13.7	\\
\hline
\end{tabular}
\\
$^*$\ epoch interval presumed to be 20 yr as no specific observation date available for epoch 1(observed between 1987 and 1989)\\
Distance reference codes: p - distances calculated from the parallax recorded in \cite{van07}.\\
s - distances calculated spectroscopically \citep{hun06}. \\
b - \citet{dip94},
c - \citet{hoe93},
d - \citet{con90}
\end{center}
\end{table*}

\subsection{Sightlines that show time variation in stellar or circumstellar spectra}

Several of our sightlines show variations in the stellar spectra or circumstellar environment. Four such stars (HD~125823, HD~137509, HD~199728 and HD~22364) are classified as magnetic chemically peculiar stars \citep{koc06}, while HD~145792 is classified as a chemically peculiar, helium strong star \citep{cat08}, and so variations 
in the spectra between epochs are to be expected. Some of the other stars have 
been grouped by their spectral classification below, explaining their spectral variations.

\subsubsection{Binary systems HD~123515 and HD~47116}

\begin{figure}
   \centering
   \includegraphics[width=\columnwidth]{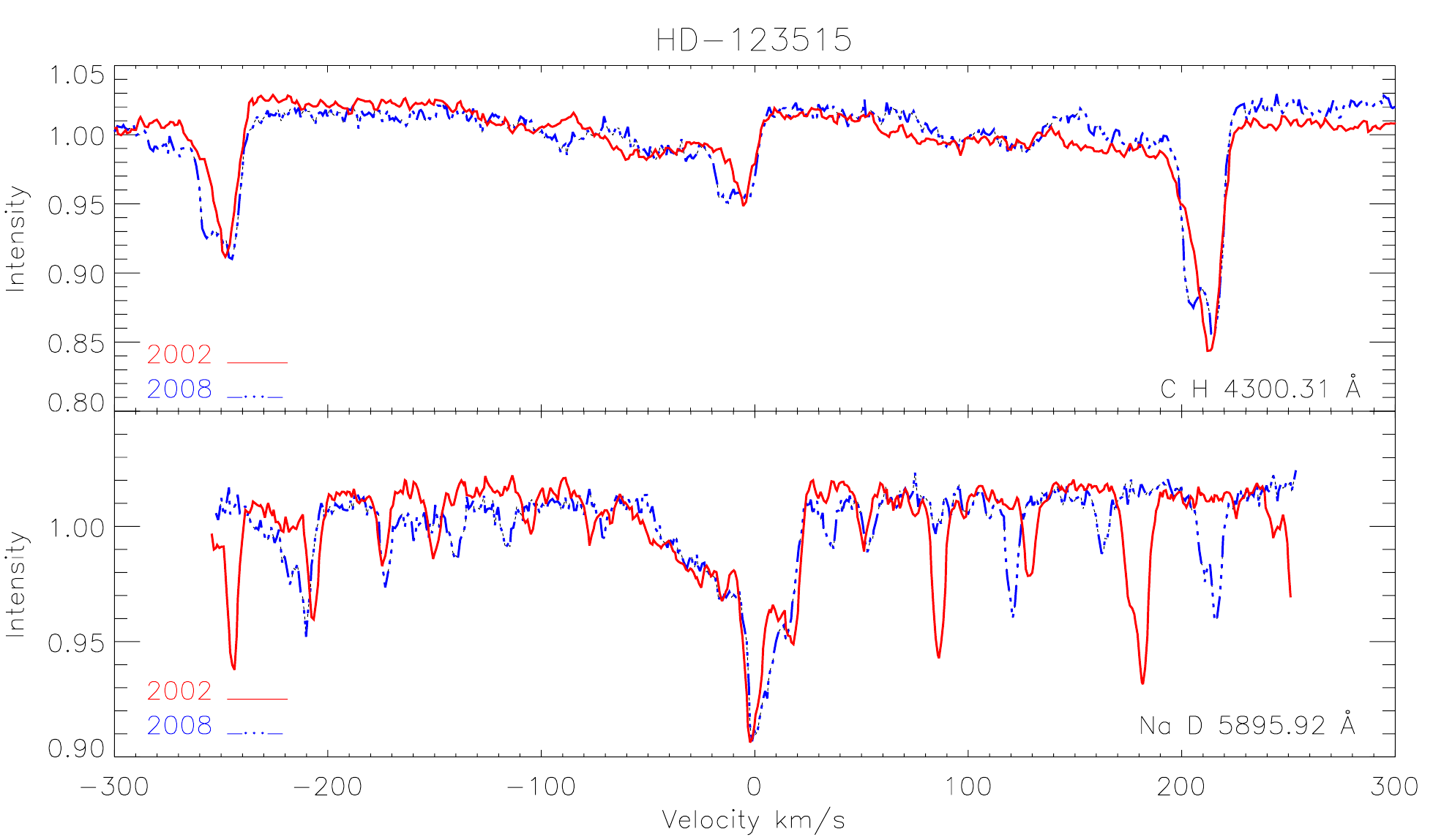}
   \caption{Overlay of epoch 1 (solid red line) and epoch 2 (dashed blue line) spectra for HD~123515, showing the areas of spectrum in the range of the CH (above) and Na~D$_2$ lines (below). The line profiles and varations visible are stellar lines varying due to stellar binarity, making interstellar variation almost impossible to detect. Many telluric lines are also seen in the Na~D$_2$ spectra.} 
   \label{f_HD123515}
\end{figure}

\begin{figure}
   \centering
   \includegraphics[width=\columnwidth]{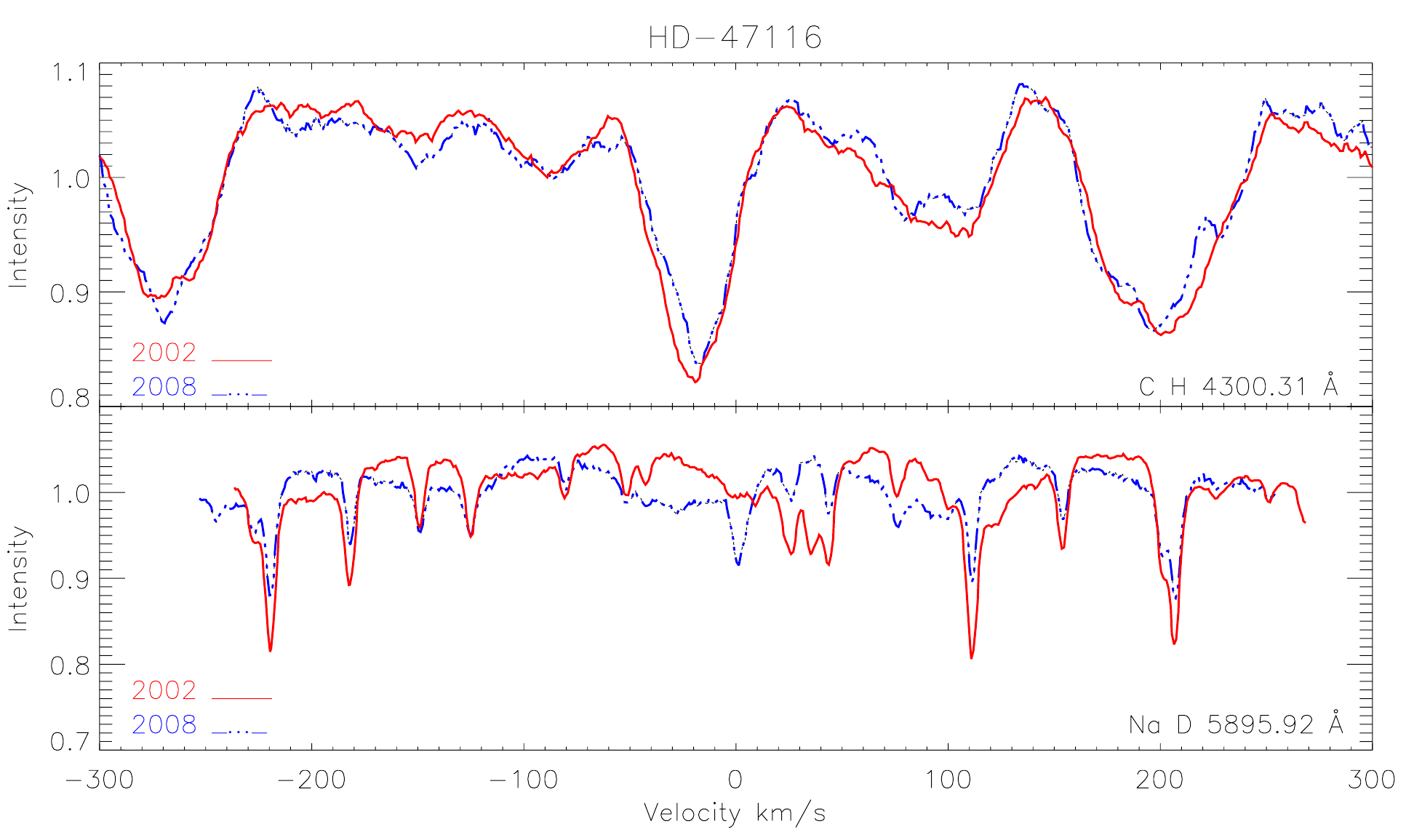}
   \caption{Overlay of epoch 1 (solid red line) and epoch 2 (dashed blue line) spectra for HD~47116, showing variations in line profile of the CH (above) and Na~D$_2$ lines (below), due to stellar binarity.} 
   \label{f_HD47116}
\end{figure}

HD~123515 is a short-period double-lined spectroscopic binary pulsating star \citep{aer99}. Figure \ref{f_HD123515} shows 
CH (4300\AA) and Na\,{\sc i} D$_2$ (5895\AA) spectra towards the object at the two epochs. HD~ 47116 
is also a star in a double system as can be seen from its spectra in Fig. \ref{f_HD47116}.

\subsubsection{B(e) stars HD~58978, HD~83312, HD~90177, HD~94910 and HD~148184}

\begin{figure}
   \centering
   \includegraphics[width=\columnwidth]{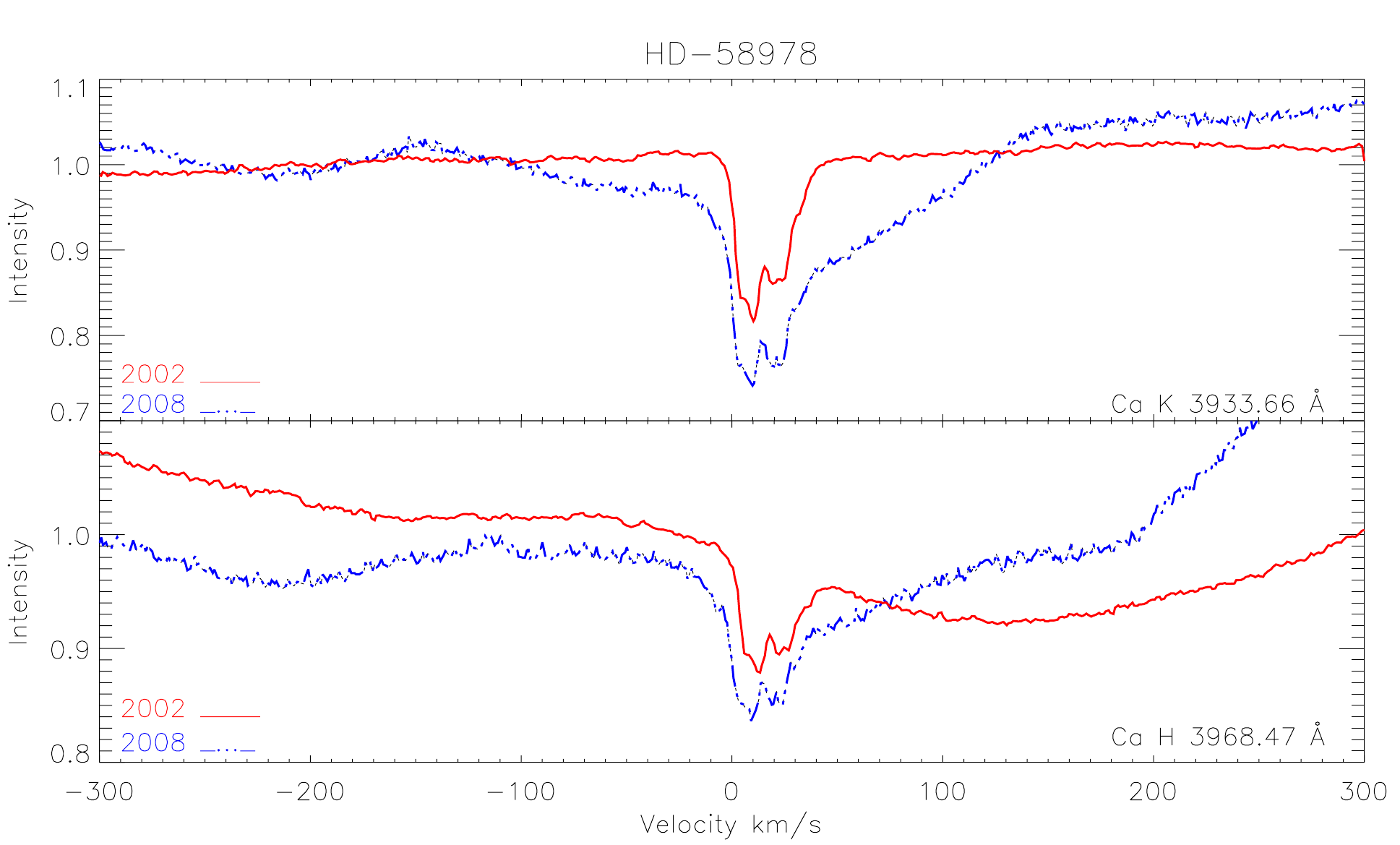}
   \caption{Overlay of epoch 1 (solid red line) and epoch 2 (dashed blue line) spectra for HD~58978, showing variations in line profile of the Ca{\sc ii} K (above) and Ca{\sc ii} H lines (below) due to B(e) star variability.} 
   \label{f_HD58978}
\end{figure}

\begin{figure}
   \centering
   \includegraphics[width=\columnwidth]{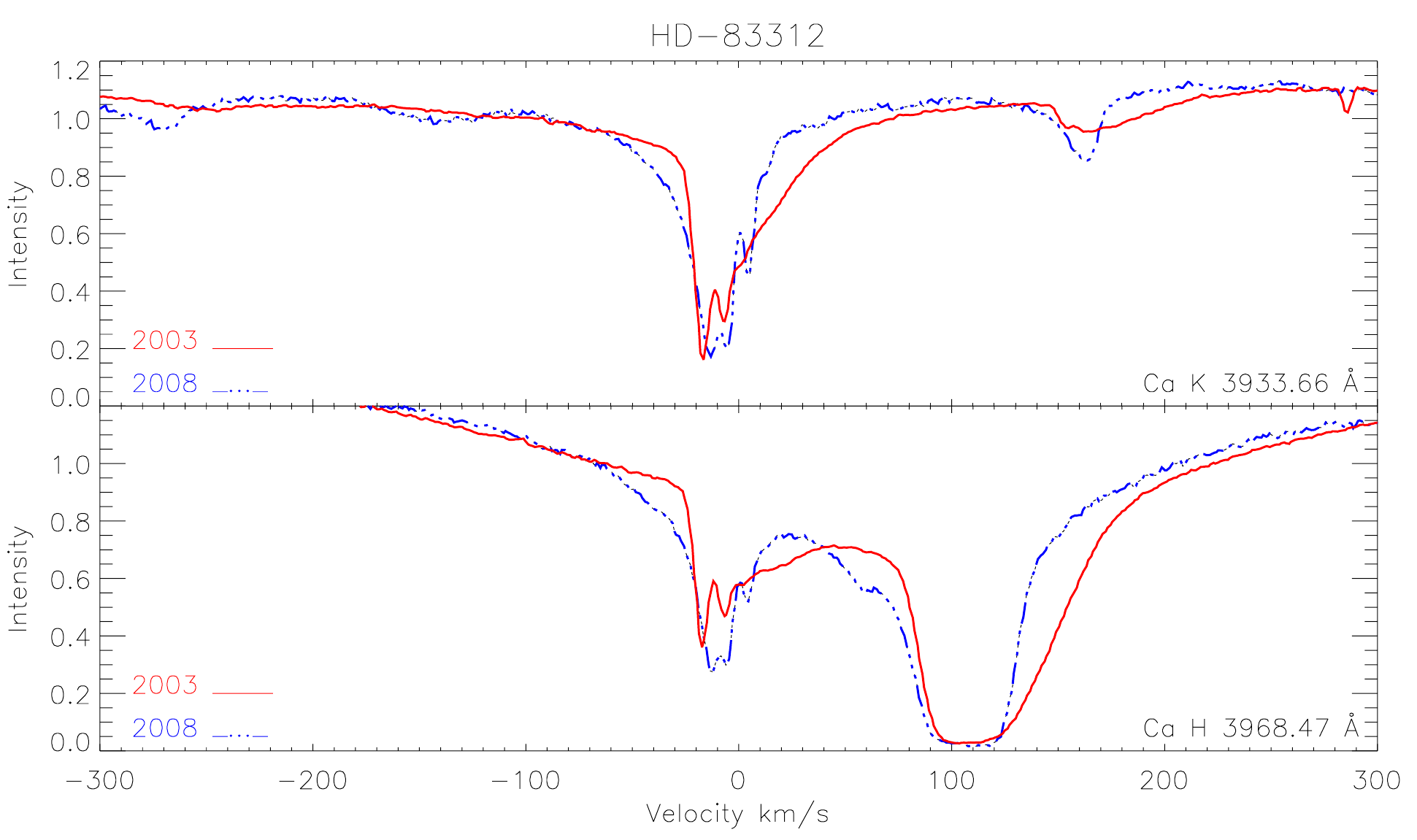}
   \caption{Overlay of epoch 1 (solid red line) and epoch 2 (dashed blue line) spectra for HD~83312, showing variations in line profile of the Ca {\sc ii} K (above) and Ca {\sc ii} H lines (below) due to B(e) star variability.} 
   \label{f_HD83312}
\end{figure}

These five objects, whose spectra show significant variation, are in fact B(e) stars.
Two examples are HD~58978 and HD~83312 whose twin-epoch spectra are shown in Figs. \ref{f_HD58978} 
and \ref{f_HD83312}, respectively. HD~58978 (FY~CMa) is a well-studied object and was found to be a B(e)+sdO binary with a period of 37.26$\pm$0.03 days by \cite{riv04}. It shows 
strong shell phases and ionization in the outer parts of the circumstellar disc. Less 
studied is the emission-line star HD~83312 \citep{hen76} which is part of the catalogue 
of young runaway $Hipparcos$ stars within 3 kpc of the Sun \citep{tet11}.     
The spectra of a typical Be star often shows clumpy 
winds with discrete accelerating components \citep[][and references therein]{ste12}. 
HD~94910, HD~90177 and HD~148184 have also been widely studied, \citep[e.g.][]{bur52,hen52}
and show the similar spectral characteristics of B(e) stars.
Any ISM lines have profiles so compromised by the stellar variation in such sightlines, that we therefore classify the changes in these spectra as circumstellar in nature, and their ISM lines are not discussed further.

\subsection{Possible time-variable spectra in the interstellar medium}

In total fix targets (6\%) from the UVES sample show apparent temporal variations in interstellar lines, which  are unlikely to be explained by stellar changes. A summary of all the sightlines where possible variation is seen can be found in Table \ref{t_column_densities_var}, and are discussed individually below.Although it is possible that some variations may be caused by slight resolution differences between epochs (see Sect. \ref{s_columndensity}), had this been the case it would be expected that variation would be much more frequently observed throughout the data set. As 80 \% show no variation we are convinced that these five sightlines show genuine evidence of fine-scale structure. 

\subsubsection{HD~112842 -- possible Na\,{\sc i} D and Ca\,{\sc ii} H \& K variations}

\begin{figure}
   \centering
   \includegraphics[width=\columnwidth]{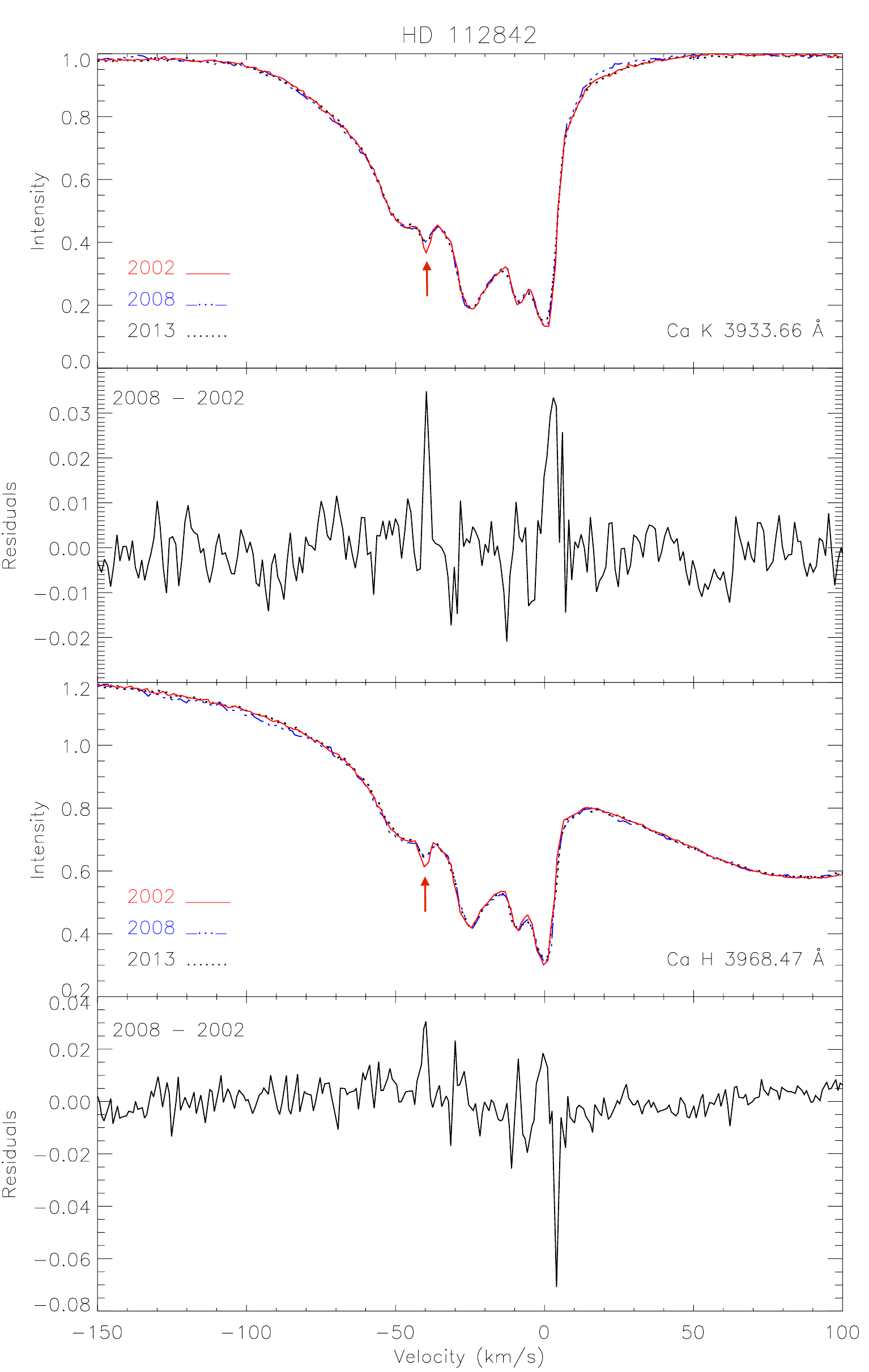}
   \caption{Overlay of epoch 1, 2002 (solid red line) and epoch 2, 2008 ( dash-dotted blue line) and epoch 3, 2013 (dotted black line) spectra for HD~112842, showing variations at $-40$\kms in the profiles of the Ca\,{\sc ii} K (top plots) and Ca\,{\sc ii} H (bottome plots) lines. The residuals between epochs 1 and 2 are plotted below each line.} 
   \label{f_HD112842_Ca}
\end{figure}

\begin{figure}
   \centering
   \includegraphics[width=\columnwidth]{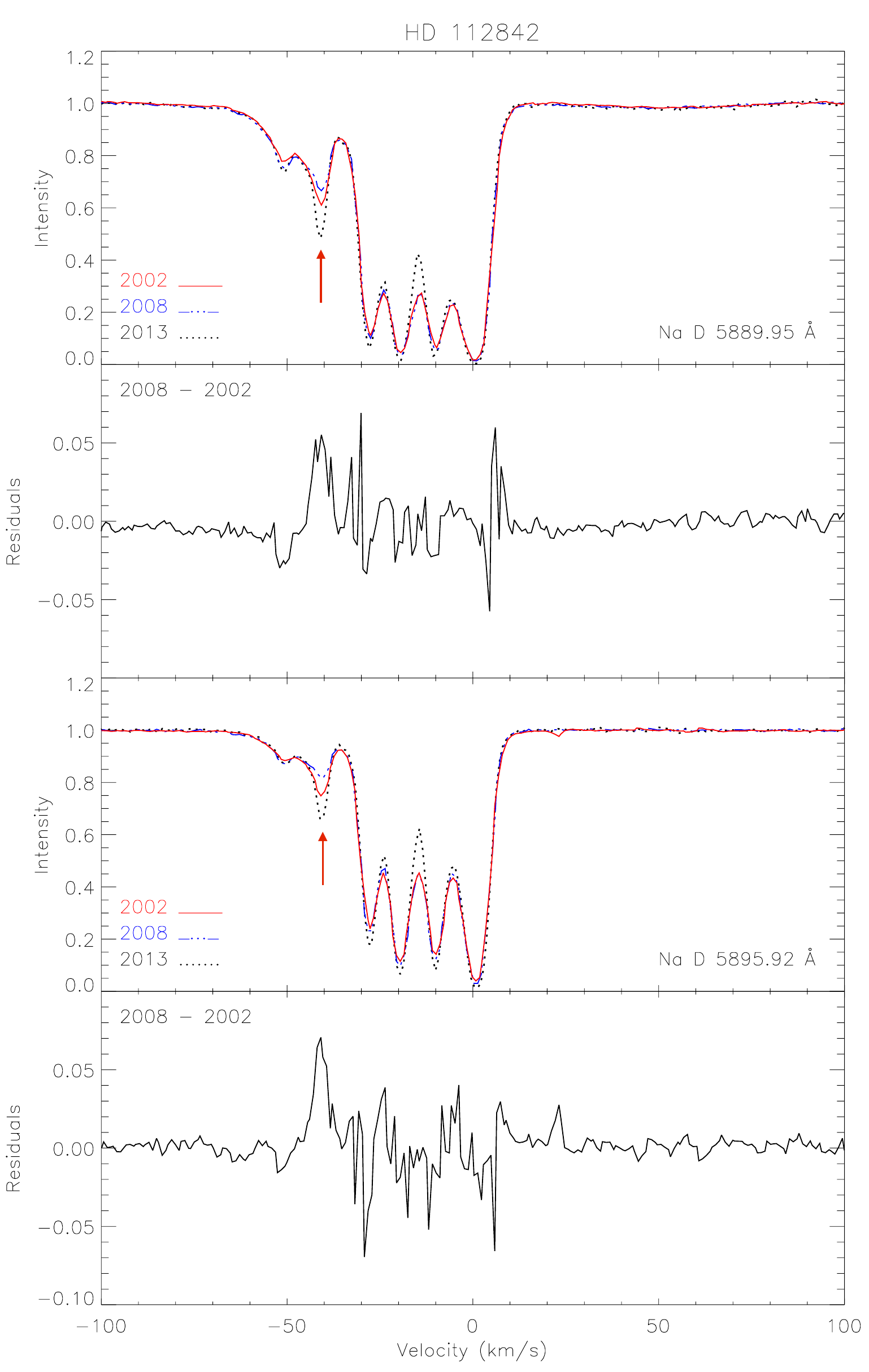}
   \caption{Overlay of epoch 1, 2002 (solid red line) and epoch 2, 2008 ( dash-dotted blue line) and epoch 3, 2013 (dotted black line) spectra for HD~112842, showing variations at $-40$\kms in the profiles of the Na~D lines (D$_2$ above and D$_1$ below). The residuals between epochs 1 and 2 are plotted below each line. Telluric lines have been removed.} 
   \label{f_HD112842_Na}
\end{figure}

Possible variations can be seen in intermediate velocity gas in the profiles of both the Na\,{\sc i} D and in Ca\,{\sc ii} H \& K lines in this sight line (Fig. \ref{f_HD112842_Ca} and Fig. \ref{f_HD112842_Na}), with the components at $-40$\kms\ 
stronger in epoch 1 in all cases.
The proper motion of this star is 9.2~au~yr$^{-1}$, and there are 5.2 yr between epochs 1 and 2. Hence the star has travelled $\sim$48~au perpendicular to the line of sight between epochs. The column density of the $-40$\kms\ components decrease from 
$\log N(\text{Na {\sc i} ({\,cm}$^{-2}$)})$=11.45 $\pm$ 0.01 to 11.36 $\pm$ 0.02, $\log N(\text{Ca {\sc ii} ({\,cm}$^{-2}$)})$=11.72 $\pm$ 0.12 to 11.33 $\pm$ 0.11, or a change of 19 percent in Na I and a factor of 2.5 in Ca {\sc ii}.  No variations are seen in other species.
For this sightline we have obtained 3rd epoch observations on 29 November 2013, which are also plotted in Fig. \ref{f_HD112842_Ca} and Fig. \ref{f_HD112842_Na}. It can clearly be seen in Ca\,{\sc ii} H \& K that there is no variation at all in the 5.6 yr and 51.9 au between epoch 2 and 3, and so residuals plotted are between epoch 1 and 2. For Na\,{\sc i} D, resolution changes are visible between 2008 and 2013, making it impossible to quantify if there is any real variation in the 3rd epoch, and so the residuals plotted are once again, only between the earlier two epochs. To verify that the variation is not solely due to differences in the resolution between epochs, Gaussian smoothing was applied to the data in an effort to make the effective instrumental resolution the same in all epochs. It was found that changes in resolution could not force all components to match in all epochs, implying that the variation is real. Ca {\sc i} may also show a decrease between epoch 1 and 2, although the quality of the epoch 2 data is not optimal at this wavelength to see convincing variation and the Ca {\sc i} lines are weak.

HD~112842 is at a distance of $\sim$2.1$\pm$0.5 kpc and lies in the Centaurus OB1 stellar association. Observations of H\,{\sc i} towards this region have revealed a large expanding shell (GSH~305+04-26) with a velocity range of $-33$ to $-17$\kms\, and estimated distance of 2.5$\pm$0.9 kpc and dimensions of 440 $\times$ 270 pc \citep{cor12}. As HD~112842 is predicted to be closer than the H\,{\sc i } shell, this implies there is fine-scale structure in the normal diffuse ISM. However, the errors on the distances are large and so there remains the possibility that the variation seen is caused by part of the shell. Na\,{\sc i} absorption is detected in the velocity range, although the variable component is more blue-shifted than expected for the shell.

\subsubsection{HD~113904 - Possible Na\,{\sc i} D variations}
HD~113904 ($\theta$~Mus) is a WR star with a binary companion (WR 48 WC5+O6/7V) and possibly has an additional O9.5/B0Iab-type supergiant companion \citep{sug08}, located in the same area of the 
sky as the Centaurus OB1 association. However the proper motion of the star is in the opposite direction to the members of this
association. Hence, \cite{cor12} deem it is unlikely to be part of Cen~OB1. \citet{stu10} give the distance as between 0.9 and 1.3 kpc, and from H$\alpha$ imaging postulate that 
the nebulosity seen near to the star is not a WR shell but in fact a possible optical supernova remnant and complex of H\,{\sc ii} regions. 
 
\begin{figure}
   \centering
   \includegraphics[width=\columnwidth]{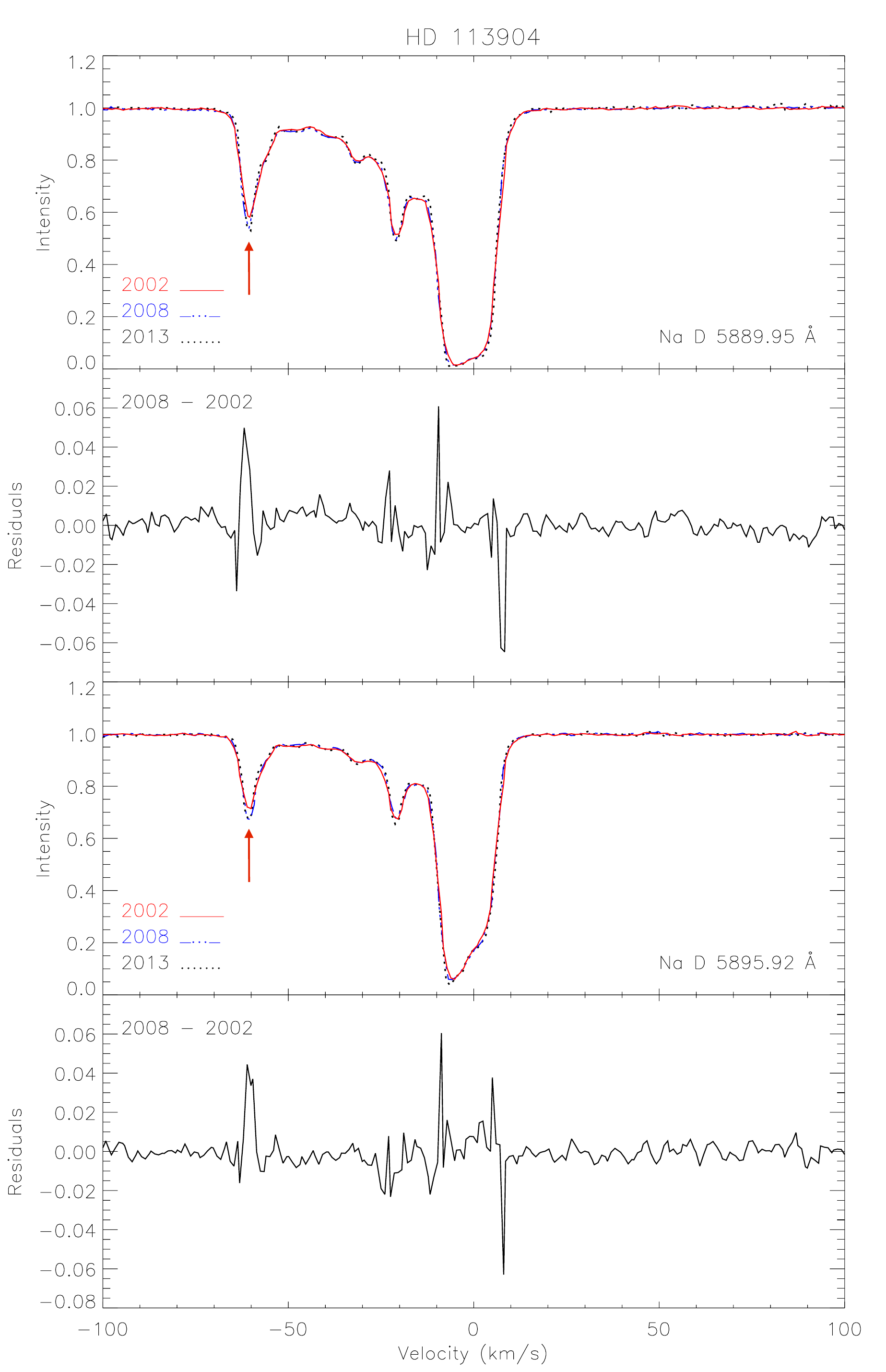}
   \caption{Overlay of epoch 1, 2002 (solid red line) and epoch 2, 2008 ( dash-dotted blue line) and epoch 3, 2013 (dotted black line) spectra for HD~113904, showing variations at $-60$\kms in the profiles of the Na~D lines (D$_2$ above and D$_1$ below). The residuals between epochs 1 and 2 are plotted below each line. Telluric lines have been removed.} 
   \label{f_HD113904}
\end{figure}

The bluest-shifted component, (see Fig. \ref{f_HD113904}), at $-60$\kms, shows { possible} variations in { intermediate velocity gas} in both Na\,{\sc i} lines in this sightline, with stronger absorption at epoch 2. As the proper motion of this star is 13.48~au\,yr$^{-1}$, and there are 6.7 yr between observations, the transverse displacement is $\sim$91~au. Although these variations are small -- $\log N(\text{Na{\sc i}})$ for this component increases from 
11.52 $\pm$ 0.01 dex to 11.56 $\pm$ 0.01 dex (10 percent) -- they can be seen in the residual plots.
For this sightline, we have obtained 3rd epoch observations from 29 November, 2013 which are also plotted in Fig. \ref{f_HD113904}. No variation is visible over the 75.8 au traversed between 2008 and 2013, so the residuals plotted are only between the earlier two epochs. There are some instrumental effects acting upon these spectra, as we can see the resolution is not identical for all three epochs. However, the variation we do see is strong enough to be identified despite these slight differences in resolution. 

\subsubsection{HD~163758 - Possible Na\,{\sc i} D variations}
\begin{figure}
   \centering
   \includegraphics[width=\columnwidth]{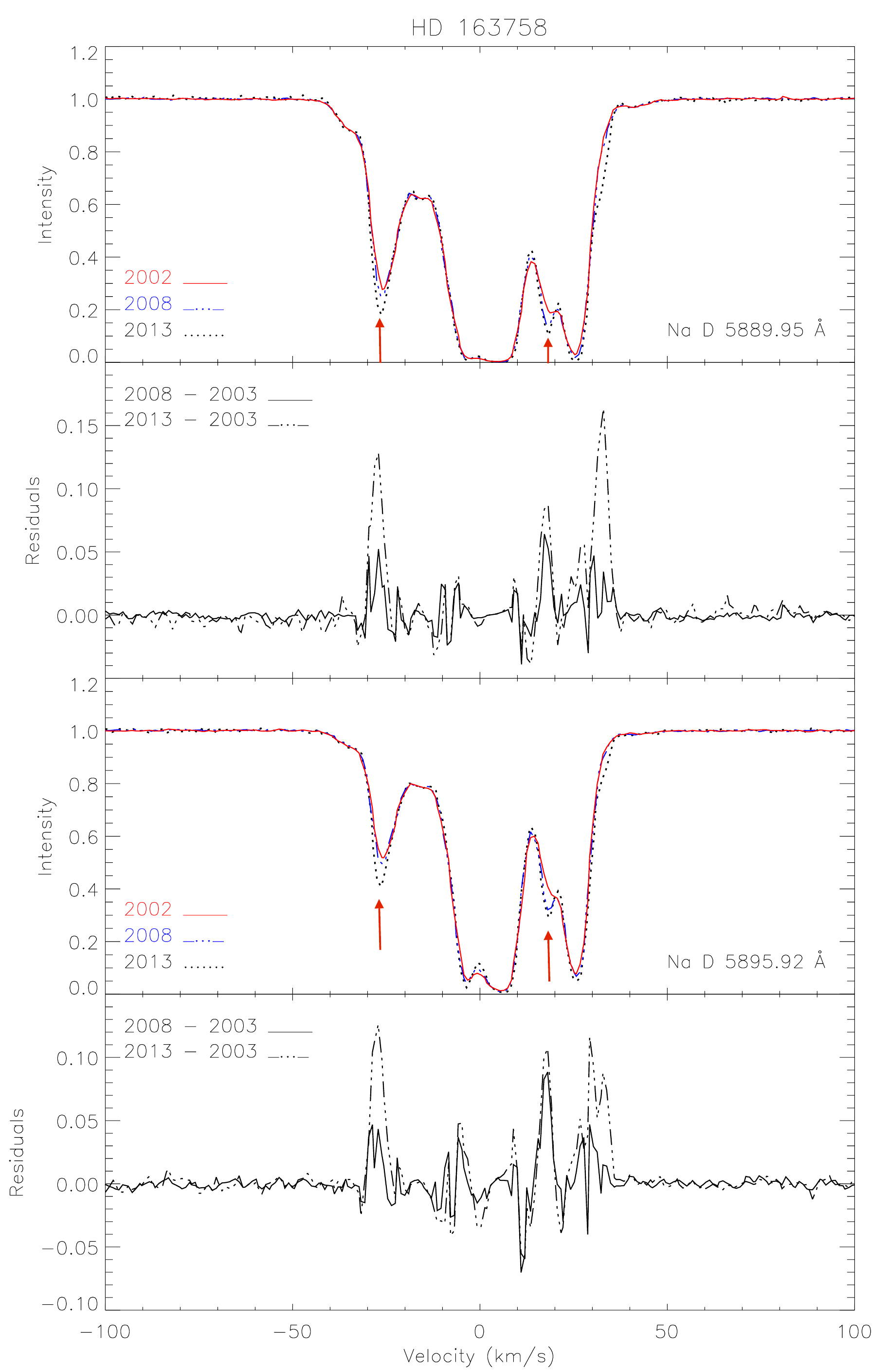}
   \caption{Overlay of epoch 1, 2002 (solid red line) and epoch 2, 2008 ( dash-dotted blue line) and epoch 3, 2013 (dotted black line) spectra for  HD~163758, showing variations at $-26$\kms\ and $19$\kms\  in the profiles of the Na~D lines (D$_2$ above and D$_1$ below). The residuals between epochs are plotted below each line.} 
    \label{f_HD163758}
\end{figure}

HD~163758 is a WR star of spectral type O6.5Iaf at a distance of 4.1 $\pm$ 1.2 kpc. For this sightline we were able to acquire epoch 3 data, observed on 19 September 2013.  In Fig. \ref{f_HD163758} we can clearly see variations in two of the components of both Na\,{\sc i} D lines, at $-26$\kms\, and $+19$\kms\, between all three epochs. Both components strengthen with time, that at +19\kms showing the largest increase from 2002 to 2008 ($\log N(\text{Na ({\,cm}$^{-2}$)})$={ 12.08 $\pm$ 2.0  to 12.63 $\pm$ 2.0, a change of a factor of 3.55)} with very little variation between 2008 and 2013. { The particularly small $b$-values { (of $\approx$ 0.5 \kms)} of these components, mean they are unresolved and therefore uncertain, leading to large errors in the column densities, which should be regarded as approximate estimates.} The component at $-26$\kms \, strengthens by a small amount between 2002 and 2008, $\log N(\text{Na ({\,cm}$^{-2}$)})$=11.87 $\pm$ 0.02 to 11.90 $\pm$ 0.01, a 7\% change, and then by a much larger degree between 2008 and 2013, to 11.99 $\pm$ 0.01, a 32\% change. These variations can very clearly be seen in the residual plots of each Na\,{\sc i} D line. Between epoch 1 and 2 the star moves 160 au in 6.7 yr and in the 5.4 yr between epoch 2 and 3 travels 130 au. We note, there are some slight differences in resolution between the three epochs due to instrumental effects, as can be seen, for example at 0 \kms, possibly bringing the weaker variations into question. { As the strongly varying component is very narrow, the changes in resolution have more effect on this component. However, even varying the spectral resolutions by 10 percent cannot account for the stronger variations that we see, (see Sect. \ref{s_columndensity})}. 

\subsubsection{HD~50896 - possible Na\,{\sc i} D variations} 
\begin{figure}
   \centering
   \includegraphics[width=\columnwidth]{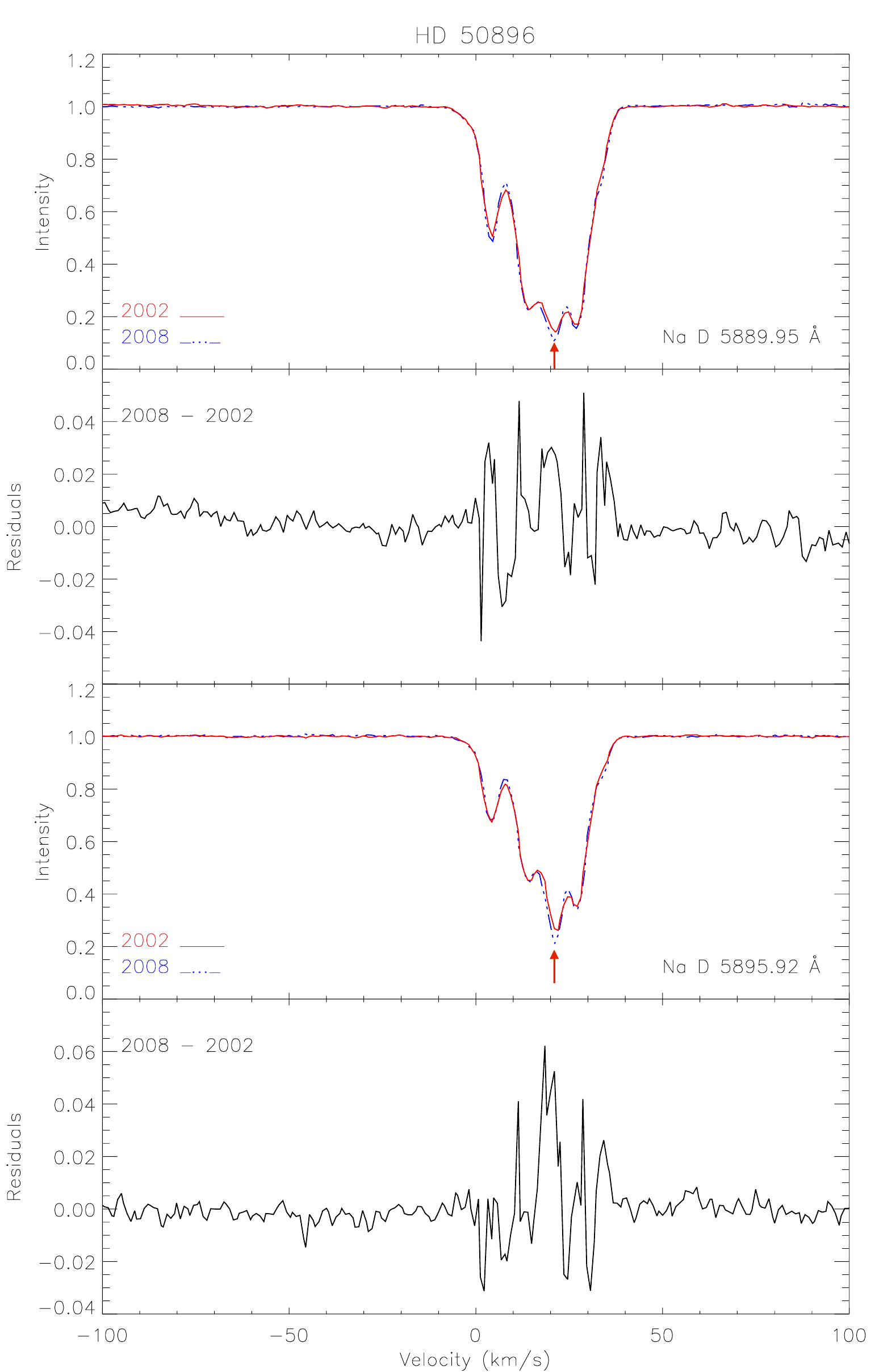}
   \caption{Overlay of epoch 1, 2002 (solid red line) and epoch 2, 2008 ( dash-dotted blue line) and epoch 3, 2013 (dotted black line) spectra for  HD~50896, showing variations at $21$\kms\ in the profiles of the Na\,{\sc i} D lines (D$_2$ above and D$_1$ below). The residuals between epochs are plotted below each line.} 
   \label{f_HD50896_Na}
\end{figure}

In the sightline of the well-studied WR star HD~50896 (WR 6) (see Fig. \ref{f_HD50896_Na}), the component at 21.3\kms\ has strengthened in both Na\,{\sc i} D lines
by a factor of { 2.3}, from $\log N(\text{Na}\ (\text{cm}^{-2}))\,={ 12.50 \pm 2.0\  \text{to}\ 
12.87 \pm 2.0} $
in epoch 2, showing { possible} variation in the 6.7 yr between the two observations. 
{ These components have particularly small $b$-values (of $\approx$ 0.5 \kms), 
meaning they are unresolved and therefore uncertain, leading to large errors in the column densities, which should be regarded as approximate estimates.} The proper motion of the star is 7.97~au~yr$^{-1}$, so the transverse displacement is only 53.2~au. These variations are { small} but tweaking the continuum fitting and resolution would only prove to increase the variations, and so we are sufficiently convinced that these variations are real.

\subsection{HD~29138 -  Ca\,{\sc i} and Ca\,{\sc ii} variations}
The variations seen in the sightline of HD~29138 are the most dramatic in the whole sample, and can very definitely not be due to differences in spectral resolution, { as changing the instrumental resolution of this sightline by 10 percent resulted in no changes in the column densities at all. }

\begin{figure}
   \centering
   \includegraphics[width=\columnwidth]{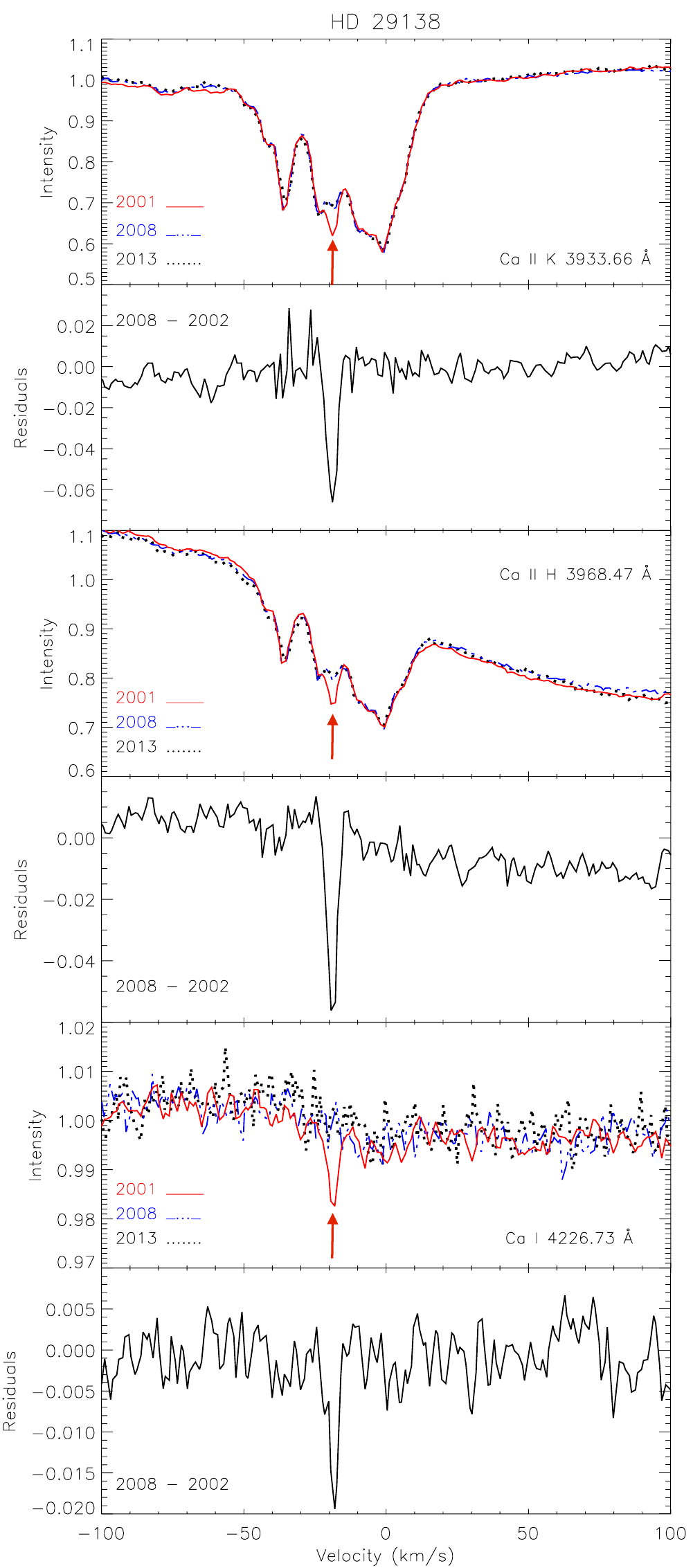}
   \caption{Overlay of epoch 1, 2002 (solid red line) and epoch 2, 2008 ( dash-dotted blue line) and epoch 3, 2013 (dotted black line) spectra for  HD~29138, showing variations at $-19$\kms\ in the profiles of calcium lines (Ca\,{\sc ii}~H above, Ca\,{\sc ii}~K middle, and Ca\,{\sc i} below). The residuals between epochs are plotted below each line.} 
   \label{f_HD29138_Ca}
\end{figure}

HD~29138 is a B1Iab star located far from the plane of the Milky Way ($b=-30.5^{\circ}$ and $d=3530$\,pc). Significant variations in Ca\,{\sc i} and both Ca\,{\sc ii} H\&K lines can be seen. All calcium lines show changes at $-18.6$\kms\ where a feature seems to diminish in all three lines in epoch 2 (Fig. \ref{f_HD29138_Ca}). The change 
in the Ca\,{\sc ii} column density of this profile is 0.13 dex (or 35 percent), from 11.53 $\pm$ 0.05 to 11.40 $\pm$ 0.01 dex. As the Ca\,{\sc i}  component disappears completely in epoch 2, we were only able to measure a column density for epoch 1. However, using the S/N of the spectra at both epochs (see Section \ref{s_no-variation}) it was possible to calculate an upper limit for the column density of Ca {\sc i} in epoch 2. The difference between this upper limit and epoch 1 is 0.47 dex, i.e. the column density decreases  from 9.60 $\pm$ 0.02 to an upper limit of 9.1 dex (change of a factor of 3.2).  
As the proper motion of this star is 32.8~au~yr$^{-1}$ these changes in the profiles occur over 210~au, due to the 6.4 yr 
between epochs. Although the changes are relatively small, the fact that we see variations in three lines and two species at the same velocity indicates that they are real. We do not see obvious variation in Na\,{\sc i} or Ti\,{\sc ii} (which is usually closely correlated with Ca\,{\sc ii}; \citep{wel97, hun06}) { which could indicate circumstellar variation \citep[c.f.][]{kie14}}. 
However, the Na D doublet is stronger than the Ca lines and so any small variation would be very difficult to see. epoch 3 data were obtained for this sightline on 19 September, 2013, 5.3 yr after epoch 2 accounting for a transversed distance of 172 au but no further variation is observed. 

The epoch 1 data also shows an unresolved line corresponding to Fe {\sc i} at 3859.9114 \AA, with central 
velocity of $-$19.2 \kms \, and column density in the optically thin approximation of $\log N$ (Fe {\sc i})$ = 11.66 $. 
The line disappears in the second epoch, with a 3$\sigma$ upper limit to its column density of $\log N$ (Fe {\sc i})$  = 11.32 $. 
The epoch 3 data are too noisy to set firm limits on the Fe {\sc i}. The Fe {\sc i} line observed in epoch 1 is close in 
velocity to that observed in Ca {\sc i} (4226 \AA) which has $v$  = -18.6 \kms.  We would expect to see a stronger Fe {\sc i} 
feature at 3719.93 \AA \, due to its larger oscillator strength, however, no absorption is detected at this 
wavelength, casting doubt on the reliability of the line observed at 3859.91 \AA. Using the Na{\sc i} D$_1$ line 
we have calculated an approximate column density for Na {\sc i} of $\log N(\text{Na ({\,cm}$^{-2}$)}) \approx$ 10.9 
dex and a Ca {\sc i} abundance of $\log N(\text{Ca ({\,cm}$^{-2}$)}) = 9.6 \pm 0.1 $ dex, yielding a relatively 
high Ca {\sc i}/Na {\sc i} ratio.  If the observed change in line strength of Fe {\sc i} is real, then the 
disappearance of both Ca {\sc i} and Fe {\sc i}, combined with the high Ca {\sc i}/Na {\sc i} ratio, suggest 
that the variable component is an example of the "CaFe" clouds noted by \citet{bon07} and discussed by \citet{gna08} 
and \citet{wel08}. The enhanced Ca {\sc i} and Fe {\sc i} in these cases could be due to some combination of 
dielectronic recombination and milder than usual depletions at temperatures above $\approx$ 5000 K.

\begin{table*}
\caption{Table of species where we see possible variation in Na {\sc i} D, Ca {\sc ii} H and K and Ca {\sc i}}
\label{t_column_densities_var}
\begin{tabular}{llllllll}  
\hline

Sightline	&	Species	&	Component &	Column density 	&	Column density	&	Column density &	time 	&	traversed 	\\

& & velocity	& (1st epoch)&  (2nd epoch) &variation (\%)	& (yr) & distance (au)\\
 \hline
HD 112842	&	Na {\sc i} D	&	-40.8 $\pm$ 0.20	&	11.45 $\pm$ 0.01	&	11.36 $\pm$ 0.02	&	19	&	5.2	&	47.8 $\pm$ 15.2 \\
	&	Ca {\sc ii} K, H	&	-39.5 $\pm$ 0.20	&	11.72 $\pm$ 0.12	&	11.33 $\pm$ 0.11	&	250	&	5.2	&	47.8 $\pm$ 15.2	\\
HD 113904	&	Na {\sc i} D	&	-60.5 $\pm$ 0.09	&	11.52 $\pm$ 0.01	&	11.56 $\pm$ 0.01	&	10	&	6.7	&	90.9 $\pm$ 28.4	\\
HD 163758	&	Na {\sc i} D	&	-25.9 $\pm$ 0.05	&	11.87 $\pm$ 0.02	&	11.90 $\pm$ 0.01	&	7	&	6.7	&	160.3 $\pm$ 58.9	\\
	&	Na {\sc i} D$^\dagger$	&	-26.0 $\pm$ 0.07	&	11.90 $\pm$ 0.01	&	11.99 $\pm$ 0.01	&	32	&	5.4	&	130.2 $\pm$ 47.8	\\
	&		Na {\sc i} D&	18.3 $\pm$ 1.00	&	12.08 $\pm$ 2.00 $^\star$	&	12.63 $\pm$ 2.00$^\star$	&	355	&	6.7	&	160.3 $\pm$ 58.9		\\
HD 50896	&	Na {\sc i} D	&	21.5 $\pm$ 0.08	&	12.50 $\pm$ 2.00$^\star$	&	12.87 $\pm$  2.00$^\star$	&	 233	&	6.7	&	53.2 $\pm$ 46.0\\
HD 29138	&	Ca {\sc i}	&	-18.6 $\pm$ 0.15	&	9.60 $\pm$ 0.02	&	9.1 ${*}$	&	320	&	6.4	&	210.2 $\pm$ 63.9	\\
	&	Ca {\sc ii} K, H	&	-18.7 $\pm$ 0.11	&	11.53 $\pm$ 0.05	&	11.40 $\pm$ 0.01	&	35 	&	6.4	&	210.2 $\pm$ 63.9	\\
\hline
\end{tabular}\\
$^\dagger$ Variation seen between epoch 2 and 3.\\
* Upper limit to column density value.\\
$\star$ Sightlines with particularly narrow $b$-values, and therefore large errors
\endcenter
\end{table*}

\subsection{Sightlines showing no time-variations}
\label{s_no-variation}

The vast majority of the sightlines in our survey show little or no variation in line profile between the two epochs for the interstellar 
species isolated and examined. However at our spectral resolution ($R$=80,000) there is always the possibility that there are blends in the absorption-lines of our observations. 

We estimated upper limits to the changes, $\Delta \log N_{Epoch 2}$, in the column density of each interstellar component by using the the expression:

\begin{equation}
\Delta \log N_{\rm epoch2} = \Delta \log N_{\rm epoch1}  \frac{(S/N_{\rm epoch1})}{(S/N_{\rm epoch2})},
\end{equation}

\noindent
where $\Delta \log N_{\rm epoch1}$ is the error in the log of the column density of each component for epoch 1 from Papers I and V, 
S/N$_{\rm epoch1}$ and S/N$_{\rm epoch2}$ are the signal to noise ratios from epoch 1 and epoch 2, respectively. 
The column density change limit has only been derived for Ca\,{\sc ii} K. 

Typical errors in the column density for epoch 1 are from 0.02 to 0.06 
 dex, being on average twice as large for epoch 2. The resulting 3$\sigma$ upper 
 limits to changes in Ca{\sc ii} are thus typically 0.1 to 0.3 dex.

None of our McDonald sightlines show very clear variation. Two of this sample, namely HD\,91316 and HD\,149757, show some possible variation. However the difference in resolution between epochs cast doubt on the reliability of such variations. As this variation is not as convincing as that of the UVES sightlines, they are not included in our final results here, but some discussion of these sightlines is provided in Appendix \ref{appendix}. 

\section{Discussion}
\label{s_dis}

In this section we discuss the main results of this survey and relate these to the idea of biased neutral formation 
in the ISM \citep{lau03}. Excluding the sightlines where we are likely seeing changes 
in absorption profiles due to circumstellar material or companions, some 1\% of the sightlines
in Ca\,{\sc i}, 2\% of those in Ca\,{\sc ii}, K and H, 
and 4\%  in Na\,{\sc i} D show evidence for time variation. None was detected in Ti\,{\sc ii}, Fe\,{\sc i}, CN, CH$^{+}$ or K\,{\sc i} (4044\AA \, line only), most probably because the 
absorption features are too weak to see changes, due to depletion or smaller $f$-values, { or in the case of Ti {\sc ii} due to a broader, smoother spatial distribution}. The smallest scale 
structure over which a change in the ISM was tentatively observed was $\sim$50~au, 
even though some 64 \% of the stars in our sample moved a smaller distance between the two epochs.  

\begin{table*}
\caption{Table of velocities, $b$-values and column densities of components towards stars where it was possible to calculate $n_e$ and lower limits for $n_H$, which are also listed in the table. { $n_H$ estimates are uncertain, see Sect. \ref{CaI/CaII} for discussion of caveats}. For sightlines where no variation was discovered, component velocities, $b$-values and column densities were taken from \citet{hun06} and Smoker \& Ledoux (in preparation)}
\label{t_column_densities}
\begin{tabular}{lrcrrlllll}  
\hline
Star 	&	Ca {\sc i}	&	&		&	Ca {\sc ii}	&	&		&	$n_e$	&	$n_H$\\
	&	v (\kms)	&	b (\kms)	&	$\log N$  ({\,cm}$^{-2}$)	&	v (\kms)	&	b (\kms)	&	$\log N$  ({\,cm}$^{-2}$)	& ({\,cm}$^{-3}$)		& ({\,cm}$^{-3}$)	\\
\hline
HD\,163758	&	- 26.7 $\pm$ 0.4 \tiny{*} &	3.4 $\pm$ 0.7	&	9.60 $\pm$ 0.10	&	-26.2 $\pm 0.1$	&	4.0 $\pm 0.1$	&	11.99 $\pm 0.01$	&	0.27 $\pm$ 0.03 \tiny{*}	&	$\ge$ 1917 $\pm$ 195  \tiny{*}\\
		&	 -4.3 $\pm$ 0.1	&	2.0 $\pm$ 0.2	&	10.13 $\pm$ 0.02	&	 -3.3 $\pm 0.1$	&	4.3 $\pm 0.2$	&	12.48 $\pm 0.02$	&	0.29 $\pm$ 0.01	&	$\ge$ 2086 $\pm$ 0.5\\
		&	4.8 $\pm$ 0.2	&	4.6 $\pm$ 0.4	&	10.13 $\pm$ 0.02	&	5.3 $\pm 0.1$	&	4.8 $\pm 0.2$	&	12.59 $\pm 0.02$	&	0.23 $\pm$ 0.01	&	$\ge$ 1647 $\pm$ 13\\
		&	24.3 $\pm$ 0.2	&	2.9 $\pm$ 0.4	&	9.78 $\pm$ 0.03	&	25.4 $\pm 0.1$	&	4.3 $\pm 0.1$	&	11.91 $\pm 0.01$	&	0.49 $\pm$ 0.03	&	$\ge$ 3497 $\pm$ 186\\							
HD\,29138 (1)	&-18.7 $\pm$ 0.1  \tiny{*}	&	1.3 $\pm$ 0.3	&	9.60 $\pm$ 0.02 	&	-18.6 $\pm$ 0.1	&	1.9 $\pm$ 0.3	& 11.50 $\pm$ 0.05	&	0.70 $\pm$ 0.10  \tiny{*}	&	$\ge$ 4515 $\pm$ 375  \tiny{*} \\
HD\,29138 (2)	&-18.7 $\pm$ 0.1  \tiny{*}	&	1.3 $\pm$ 0.3	&	$\le$ 9.10 	&	-18.5 $\pm$ 0.1	&	1.9 $^\dagger$	& 11.40 $\pm$ 0.01	&	$\le$ 0.3 \tiny{*}	&	-\\

HD\,30677	&	12.1	$\pm$	0.2	&	0.9	$\pm$	0.0	&	9.93	$\pm$	0.08	&	11.8	$\pm$	0.7	&	4.0	$\pm$	0.3	&	11.70	$\pm$	0.02	&	1.12	$\pm$	0.17	&	$\ge$ 8006	$\pm$	1186	\\
HD\,33328	&	-11.5	$\pm$	0.1	&	2.0	$\pm$	0.4	&	9.71	$\pm$	0.08	&	-11.8	$\pm$	0.7	&	4.0	$\pm$	0.1	&	11.86	$\pm$	0.01	&	0.47	$\pm$	0.08	&	$\ge$ 3337	$\pm$	584	\\
	&	0.8	$\pm$	0.5	&	2.7	$\pm$	2.7	&	9.25	$\pm$	0.30	&	0.0	$\pm$	0.4	&	2.2	$\pm$	0.1	&	11.38	$\pm$	0.01	&	0.49	$\pm$	0.46	&	$\ge$ 3495	$\pm$	3319	\\
HD\,58343	&	1.2	$\pm$	0.4	&	2.6	$\pm$	0.7	&	9.69	$\pm$	0.06	&	1.6	$\pm$	0.4	&	2.5	$\pm$	0.2	&	11.84	$\pm$	0.02	&	0.47	$\pm$	0.05	&	$\ge$ 3337	$\pm$	322	\\
HD\,76341	&	5.8	$\pm$	0.1	&	3.1	$\pm$	3.1	&	9.77	$\pm$	0.08	&	6.3	$\pm$	0.8	&	4.8	$\pm$	0.1	&	12.12	$\pm$	0.02	&	0.29	$\pm$	0.04	&	$\ge$ 2106	$\pm$	312	\\
HD\,112272	&	-17.7	$\pm$	0.4	&	3.5	$\pm$	0.7	&	9.88	$\pm$	0.08	&	-16.9	$\pm$	0.1	&	3.5	$\pm$	0.1	&	12.16	$\pm$	0.02	&	0.35	$\pm$	0.05	&	$\ge$ 2474	$\pm$	367	\\

HD\,115842	&	-14.8	$\pm$	0.5	&	0.4	$\pm$	0.4	&	9.7	$\pm$	0.08	&	-14.9	$\pm$	0.2	&	3.3	$\pm$	0.3	&	12.24	$\pm$	0.02	&	0.19	$\pm$	0.03	&	$\ge$ 1360	$\pm$	201	\\
HD\,136239	&	-52.6	$\pm$	0.2	&	2.3	$\pm$	2.3	&	9.93	$\pm$	0.08	&	-52.3	$\pm$	0.9	&	5.5	$\pm$	0.5	&	12.60	$\pm$	0.03	&	0.14	$\pm$	0.02	&	$\ge$ 1008	$\pm$	123	\\
HD\,142758	&	-43.8	$\pm$	0.1	&	1.1	$\pm$	0.0	&	9.86	$\pm$	0.08	&	-44.2	$\pm$	0.1	&	5.0	$\pm$	0.1	&	12.22	$\pm$	0.02	&	0.29	$\pm$	0.04	&	$\ge$ 2058	$\pm$	305	\\
	&	-11.7	$\pm$	0.2	&	1.4	$\pm$	1.4	&	9.54	$\pm$	0.08	&	-12.2	$\pm$	0.2	&	3.0	$\pm$	0.2	&	12.12	$\pm$	0.02	&	0.17	$\pm$	0.03	&	$\ge$ 1240	$\pm$	184	\\
	&	2.3	$\pm$	0.5	&	2.8	$\pm$	0.6	&	9.43	$\pm$	0.08	&	2.4	$\pm$	0.1	&	5.0	$\pm$	0.1	&	12.01	$\pm$	0.02	&	0.17	$\pm$	0.03	&	$\ge$ 1240	$\pm$	184	\\
HD\,148937	&	3.0	$\pm$	0.2	&	1.7	$\pm$	0.3	&	9.67	$\pm$	0.08	&	2.9	$\pm$	0.5	&	3.1	$\pm$	0.2	&	12.26	$\pm$	0.02	&	0.17	$\pm$	0.03	&	$\ge$ 1212	$\pm$	180	\\
HD\,148379	&	-18.1	$\pm$	0.5	&	5.8	$\pm$	1.2	&	10.35	$\pm$	0.08	&	-18.5	$\pm$	1.4	&	8.0	$\pm$	0.3	&	12.50	$\pm$	0.02	&	0.47	$\pm$	0.07	&	$\ge$ 3337	$\pm$	494	\\
	&	-12.1	$\pm$	0.1	&	0.4	$\pm$	0.4	&	9.96	$\pm$	0.08	&	-11.4	$\pm$	0.9	&	5.5	$\pm$	0.7	&	11.75	$\pm$	0.04	&	1.07	$\pm$	0.10	&	$\ge$ 7646	$\pm$	738	\\
HD\,148688	&	2.8	$\pm$	2.0	&	3.1	$\pm$	0.6	&	9.98	$\pm$	0.08	&	3.7	$\pm$	0.9	&	5.2	$\pm$	0.2	&	12.59	$\pm$	0.01	&	0.16	$\pm$	0.03	&	$\ge$ 1157	$\pm$	202	\\
HD\,152003	&	2.3	$\pm$	0.2	&	4.5	$\pm$	0.9	&	10.26	$\pm$	0.08	&	2	$\pm$	0.2	&	3.0	$\pm$	0.7	&	12.17	$\pm$	0.04	&	0.81	$\pm$	0.08	&	$\ge$ 5800	$\pm$	560	\\
HD\,154811	&	-19	$\pm$	0.4	&	0.8	$\pm$	0.8	&	9.21	$\pm$	0.30	&	-19.4	$\pm$	0.1	&	3.4	$\pm$	0.5	&	10.89	$\pm$	0.04	&	1.38	$\pm$	1.13	&	$\ge$ 9850	$\pm$	8074	\\
	&	6.7	$\pm$	0.2	&	3.1	$\pm$	0.6	&	9.96	$\pm$	0.08	&	6.5	$\pm$	0.6	&	3.6	$\pm$	0.1	&	12.16	$\pm$	0.02	&	0.42	$\pm$	0.06	&	$\ge$ 2975	$\pm$	440	\\
HD\,154873	&	-23.7	$\pm$	0.1	&	2.8	$\pm$	2.8	&	9.12	$\pm$	0.3	&	-23.7	$\pm$	0.1	&	5.6	$\pm$	0.2	&	11.64	$\pm$	0.02	&	0.20	$\pm$	0.18	&	$\ge$ 1424	$\pm$	1289	\\
	&	-12.8	$\pm$	0.1	&	3.8	$\pm$	0.8	&	9.95	$\pm$	0.08	&	-13	$\pm$	0.8	&	4.6	$\pm$	0.1	&	12.18	$\pm$	0.01	&	0.39	$\pm$	0.07	&	$\ge$ 2776	$\pm$	486	\\
	&	0.5	$\pm$	0.1	&	3.8	$\pm$	0.8	&	9.95	$\pm$	0.08	&	0.3	$\pm$	0.5	&	3.1	$\pm$	0.1	&	12.03	$\pm$	0.01	&	0.55	$\pm$	0.10	&	$\ge$ 3921	$\pm$	686	\\
	&	7.3	$\pm$	0.1	&	1.7	$\pm$	0.0	&	10.07	$\pm$	0.08	&	7.1	$\pm$	0.5	&	2.8	$\pm$	0.1	&	12.19	$\pm$	0.02	&	0.50	$\pm$	0.07	&	$\ge$ 3576	$\pm$	530	\\
HD\,155806	&	-36.5	$\pm$	0.4	&	2.0	$\pm$	2.0	&	9.47	$\pm$	0.30	&	-37	$\pm$	0.1	&	4.9	$\pm$	0.5	&	11.47	$\pm$	0.04	&	0.66	$\pm$	0.54	&	$\ge$ 4714	$\pm$	3864	\\
	&	-2.4	$\pm$	0.5	&	1.8	$\pm$	1.8	&	9.47	$\pm$	0.30	&	-2.2	$\pm$	0.4	&	2.2	$\pm$	0.8	&	11.52	$\pm$	0.06	&	0.59	$\pm$	0.43	&	$\ge$ 4202	$\pm$	3100	\\
HD\,163800	&	1.2	$\pm$	0.0	&	6.5	$\pm$	6.5	&	9.48	$\pm$	0.30	&	0.6	$\pm$	0.4	&	5.0	$\pm$	0.1	&	11.55	$\pm$	0.02	&	0.56	$\pm$	0.51	&	$\ge$ 4013	$\pm$	3633	\\
	&	7.0	$\pm$	0.0	&	2.1	$\pm$	0.4	&	10.39	$\pm$	0.08	&	6.3	$\pm$	0.5	&	2.8	$\pm$	0.5	&	12.44	$\pm$	0.02	&	0.59	$\pm$	0.09	&	$\ge$ 4202	$\pm$	622	\\
\hline
\end{tabular}
$\dagger$ - b value fixed\\

* - Sightlines with varying components in either Ca or Na\\

For HD 29138, the (1) and (2) indicate the epoch in which these column densities were measure, ie, (1) was in 2002 and (2) in 2008. As none of the calcium lines varied in the other sightlines between epochs, these values apply for both 2002 and 2008 data. \\
\endcenter
\end{table*}

Optical and H\,{\sc i} observations of interstellar clouds imply gas densities several orders of magnitude higher than the background ISM \citep[e.g.][]{dia89,cra02,smi13}, so they are not in pressure equilibrium with the surrounding gas. Such clouds would quickly evaporate, so must be continuously generated or confined by additional sources of pressure. \citet{hei97} invoked the idea of non-spherical geometry to solve this pressure problem. Cold and dense sheets or filaments aligned along the line of sight could explain the column density variations. \citeauthor{hei97} also suggested that observations of high proper motion stars should show variation in almost all sightlines, which is not the case in this study. Also, evidence for the low temperatures ($\sim10$\,K) required by the \citeauthor{hei97} model have not been found in the cold ISM, although \citet{jen11} have detected a small but significant fraction of the diffuse ISM which is at high pressure.

Studies of trace and dominant ions in the UV have instead indicated that the variations may be due to local differences in ionization rather than density \citep{dan01,lau03,wel07}. However, this does not easily explain directly-observed variations in H\,{\sc i} { \citep{Fra94,dia89,bro05}}, nor the existence of au-scale structure in maps of dust emission \citep{miv10}. Some authors have proposed alternative models, in which the observed variations are due to statistical fluctuations caused by a power spectrum of cloud sizes \citep{des00,des07}.

To date, only about 25 examples of temporal variation are known \citep{lau07}. Two notable searches for variation were those by \citet{dan01} who studied ten stars behind the Vela supernova remnant and detected four sightlines which varied, and \citet{smo11} who found evidence of variation towards three out of 46 stars spread across the southern sky - similar to the numbers of varying sightlines found in this study.

\subsection{Cloud electron densities and estimates of number densities and sizes}\label{CaI/CaII}
At the low densities in the ISM, ionization states are determined by the balance between photoionization and recombination. For neutral and first ionization states in equilibrium this can be expressed as;
\begin{equation}
\varGamma (X_I) n (X_I) = \alpha_r (X_I,T)n(X_{II}) n_e
\end{equation}

where $\varGamma$ is the photoionization rate, $\alpha_r$ is the radiative recombination rate coefficient, $n_e$ is the electron number density and $n(X)$ is the number density of species $X$ \citep{wel03}. 

Column densities of Ca\,{\sc i} and Ca\,{\sc ii}  can be employed to estimate the electron density of the absorbing medium if it is in Ca ionization equilibrium and at a temperature of about 100 K. If we assume a constant density along the line of sight, $n(X)$ may be replaced with $N(X)$, i.e. number density with column density, allowing us to rewrite equation (2) for the specific case of Ca\,{\sc i} and Ca\,{\sc ii} which becomes:

 \begin{equation}
n_e = \frac {\varGamma(Ca_I)}{\alpha_r (Ca_I, T)}\, \frac{N(Ca_I)}{N(Ca_{II})}
\end{equation}

It is necessary to assume that the Ca\,{\sc i} and Ca\,{\sc ii} lines sample the same material. In the sightlines where the $b$-value for Ca\,{\sc i} is smaller than that of Ca\,{\sc ii}, it is more likely that the former lies in denser regions, due to enhanced recombination. Hence, in the following discussion, it is important to bear in mind the caveat that Ca\,{\sc i} and Ca\,{\sc ii} may trace different regions \citep[See][for further discussion]{cra02}.
  
Using the methods of \citet{wel03} and \citet{smi13} we have estimated the electron density of components in those sightlines where we see variation, and also for a sample without (where Ca\,{\sc i} and Ca\,{\sc ii} were present). These values are listed in table \ref{t_column_densities}. The sample was chosen using the criterion that both Ca {\sc ii} and Ca {\sc i} components found in \citet{hun06} and Smoker \& Ledoux (in preparation) had comparable velocities (within 1\kms of each other). As these values can only be calculated when Ca\,{\sc i} is visible, Table \ref{t_column_densities} is therefore heavily biased towards the densest material, where Ca I is most abundant and thus easiest to detect.

Assuming that the electrons released through photoionization of atomic carbon dominate the electron density \citep{cra02}, we can use this to estimate a lower limit to the total number density of the sightline. Adopting a carbon depletion of 60\% implies a gas-phase carbon abundance of 1.4 $\times$ 10$^{-4} n_H$ \citep{sof97}. Values of $n_H$ for each sightline where we have calculated $n_e$ are shown in Table \ref{t_column_densities}. These density estimates (at least those with $n_H \approx>10^3$ cm$^{-3}$) are consistent with the filament structure predicted by \cite{hei97}, as found in \cite{cra02}. Typical values of $n_H$ are 10 - 100, 500, and 500-5000 cm$^{-3}$ for the diffuse atomic, diffuse molecular and translucent ISM respectively \citep{sno06}. As can be seen from Table \ref{t_column_densities}, all of our $n_H$ values lie in or beyond the upper range. Typical values stated in \cite{wel07} of $n_e$ for the diffuse neutral gas are $\approx$ 0.05-0.20 cm$^{-3}$, calculated from C {\sc i}, Na {\sc i} and/or K {\sc i}. Again all but five of the densities found in Table \ref{t_column_densities} are larger than this "normal" range. Although the values of $n_e$ and $n_H$ are high for the components in which we detect variations, it is interesting to note that they are in agreement with those for components that do not show variation. 

It has generally been found that electron densities estimated from calcium tend to be higher than those derived from other species observed in the same sightline \cite[e.g.][]{wel99, son03}. \cite{wel03} have discussed this effect in detail. The cause of this effect is unknown \citep{wel03}, but it has only been reported towards a handful of stars. In the absence of other tracers we cannot exclude this possibility. It is also possible that the derived densities are underestimated due to the aforementioned caveat that Ca\,{\sc i} and Ca\,{\sc ii} may trace different regions. If Ca\,{\sc ii} is mostly from an outer envelope, and Ca\,{\sc i} from a more dense region as discussed in \cite{cra02}, then the Ca\,{\sc i}/Ca\,{\sc ii} ratio (and hence the density) will be greater than the value listed in Table \ref{t_column_densities}. 

We were able to calculate densities for 32 velocity components towards 19 stars, of which two components were found to include time-variable species (Sect. 3). High number densities of $>10^3$ cm$^{-3}$ are derived for all of the components. It appears that at least some regions of the diffuse ISM are at densities which are much higher than the typical values of $10^1-10^2$ cm$^{-3}$. If the assumptions leading to these values are correct, they lend credence to the high densities of $(1-2) \times 10^4$ cm$^{-3}$ derived by \cite{cra02} and Smith et al (2013) for $\kappa$~Vel. It is also consistent with the detection of a small but significant population of diffuse ISM components which are at much higher pressure than the surround medium \citep{jen11}. { Conversely, density estimates from C {\sc i} by \citet{jen11} are usually significantly lower than those estimated from Ca{\sc i}/Ca{\sc ii} (Welty et al. in preparation).}

\subsection{Biased Neutral formation in the ISM}
As in previous ISM studies [see][for a review]citep{lau07}, changes in Na\,{\sc i} are more often observed than 
in Ca\,{\sc ii}. \citet{lau03} estimate the minimum apparent length for the size of fluctuations 
observable in different species, being set by the distance an atom can travel before it is ionised, i.e. the 
velocity divided by the photoionization rate. Using reasonable values for the velocity (estimated from the $b$-values) 
and photoionization rate, \citet{lau03} estimated that Na\,{\sc i} has an apparent scale of around 250~au, five times smaller than predicted for Ca\,{\sc ii}, hence explaining why typically tiny-scale structure is 
more often seen in Na\,{\sc i}. However, due to the higher photoionization rate of Ca\,{\sc i}, this species should 
show larger time-variability than Na\,{\sc i}. On the other hand, the relative weakness of the Ca\,{\sc i} line means this 
is hard to test in practice, although we have attempted to see if such variation would be detected in the 
Ca\,{\sc i} spectra where there is corresponding time-variability in Na\,{\sc i}.

In the current sample there are five sightlines for which we have multi-epoch data for Ca\,{\sc i}. 
Of these, one shows variability in both Ca\,{\sc i} and Ca\,{\sc ii} (HD\,29138) and three
either have no observed Ca\,{\sc i} (HD\,50896 and HD\,112842), or have S/N insufficiently 
high to have detected the same percentage change in column density as seen in Na\,{\sc i} (HD\,113904). 
The remaining target where { time variability} 
is seen in Na\,{\sc i} (HD\,163758) shows no Ca\,{\sc i} 
variation, and hence provides evidence against the biased neutral formation model. However, we caution
that this star is WR, so there is always the possibility that we are observing
changes in the circumstellar environment. Our results are therefore inconclusive in proving or 
disproving the biased neutral formation in the interstellar gas model of \citet{lau03}. 

\section{Summary}
\label{s_summary}

We have observed 104 early-type stars at twin epochs to search for time-variable structures in the 
ISM over periods of 5 to 20 yr. Our data set included 83 spectra with $R \simeq$ 80,000 from UVES and 
21 spectra with $R \simeq$ 140,000 from the 2.7m telescope at McDonald Observatory. Excluding changes in absorption likely caused by eclipsing 
binaries or circumstellar discs, we find 5 cases within the UVES data of possible variation in the ISM. 
We have obtained follow-up third-epoch UVES and McDonald observations for the few objects which 
show possible change in their interstellar profiles. In only one case do we detect further variation. 
We expect this lack of variation, compared to \citet{wat96} who found differences in Na {\sc i} towards the individual members of all 17 of the binary or multiple star systems they observed (with separations 480-29000 au), is due to the much smaller scales probed in our investigation, i.e. from $\approx$ 0 -- 360 au. Hence, although fine-scale structure seems ubiquitous on scale $\ga$ 480 au it does not seem to be the case on scales $\la$ 360 au.  
We were able to calculate { approximate} densities, { derived from the ratio of Ca{\sc i} and Ca{\sc ii}}, in 32 velocity components towards 19 stars, of which two components were found to include time-variable species (section 3 and 4). Density estimates are consistent with the filament structure predicted by \cite{hei97}, as more recently found in \cite{cra02}. The majority of electron densities are higher than the typical values of 0.05 -- 0.20 cm$^{-3}$, { derived from other trace-dominant ratios} \citep{wel07}. The components in which we have found variation show no significant difference in density calculations to those where we found no variation. 
Future work will include FEROS and HARPS data taken from the ESO archive to study an even larger sample of objects 
and investigate tiny-scale structure in terms of the power spectrum of the ISM.

\section{Acknowledgements}
Based on observations made with ESO telescopes at the La Silla Paranal Observatory under programme IDs 266.D-5655(A), 081.C-0475(A) 
and 092.C-0173(A) and data taken at The McDonald Observatory of The University of Texas at Austin, for which we gratefully thank 
the support staff including David Doss. We thank Thomas Rivinius for advice on Be star spectra. This work was supported by the 
visiting research fund of Queen's University Belfast and the ESO Director General's Discretionary Fund. 
CMM is grateful to the Department of Education and Learning (DEL) in Northern Ireland and Queen's University Belfast for the 
award of a research studentship. This research has made use of the SIMBAD database,
operated at CDS, Strasbourg, France \citep{wen00}. Thanks to an anonymous referee for useful comments on the initial version of the manuscript

\def\apj{ApJ}
\def\apjl{ApJL}
\def\apjs{ApJS}
\def\aap{A\&A}
\def\aaps{A\&AS}
\def\aapr{A\&ARv}
\def\araa{ARA\&A}
\def\jrasc{JRASC}
\def\pasj{PASJ}
\def\aj{AJ}
\def\nat{Nature}
\def\mnras{MNRAS}
\def\pasp{PASP}
\def\ssr{Space Sci. Rev.}
\def\actaa{Acta Astron.}

\bibliographystyle{mn2e}
\bibliography{reference}

\appendix

\section{Possible Variation in two McDonald stars} \label{appendix}
\subsection{HD~149757  Na\,{\sc i} D variations}

HD~149757 ($\zeta$~Oph) is a runaway O9V B(e) star at a distance of $\sim$200~pc. The interstellar absorption-lines towards this star are very well studied, and the sightline often serves as an exemplar for modelling the ISM. Any variations in interstellar absorption towards this star are therefore potentially very important. Its displacement between the two epochs is 115.5~au.

\begin{figure}
   \centering
   \includegraphics[width=\columnwidth]{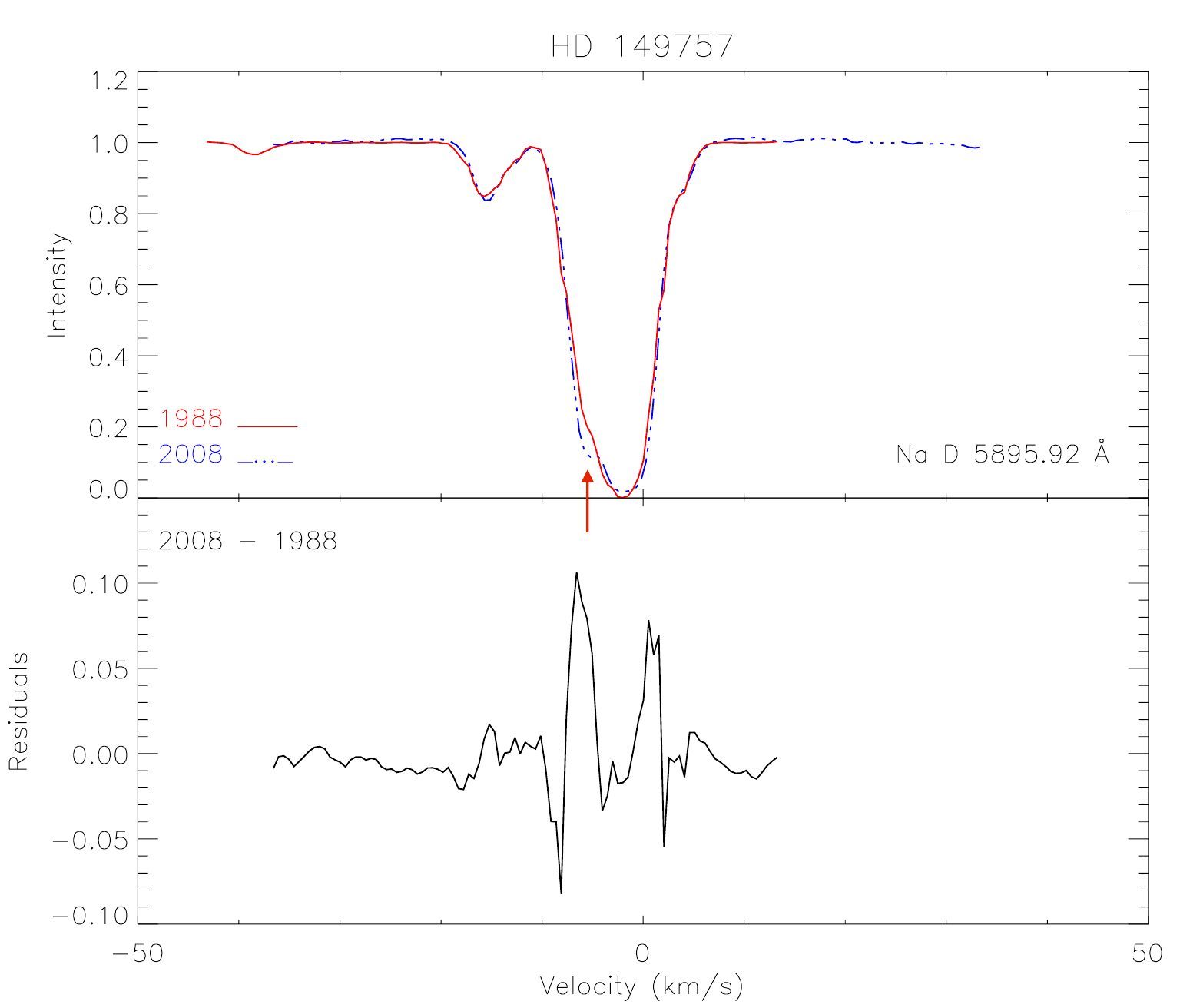}
   \caption{Overlay of WHK spectra (solid red line) and McDonald data (dashed blue line line) for HD~149757 ($\zeta$~Oph) showing variations at $-5.6$\kms\ in the line profile for the Na\,{\sc i} D line at 5895 \AA. The residuals between epochs are plotted below. It was necessary to degrade the resolution of the data used by WHK.} 
   \label{f_HD149757}
\end{figure}

The Na\,{\sc i} D absorption component at $-5.6$\kms\ shows a possible change between the two epochs separated by 20 yr in D$_1$. This variation is exaggerated when the McDonald data are compared with the models. However, when the original data (only available for Na D 5895.92~\AA, see Fig. \ref{f_HD149757} ) were smoothed to allow the resolution to match that of the epoch 2 McDonald observations, the variation was less obvious. This implies a significant amount of the variation detected was due to the difference in resolution. Nonetheless this variation is still seen (however weak) in the comparison with the epoch 1 data. { This sightline has been studied in detail by \citet{bar95} and so we have also produced a fit using the velocity and column density values taken from \citet{bar95}, but with the instrumental $b$-value from the McDonald observations replacing those of UHRF. The resulting fit also indicates that the component at $-5.6$\kms shows some possible variation. }

To investigate this variation further, we searched the ESO archives and found four observations with HARPS of Na\,{\sc i} D towards this star, taken between 2006 and 2014. No further variations were detected, but from 2006 onwards the $-5.6$\kms\ absorption is saturated and so continued increases in column density might not be visible. Future investigation of this component will need to examine the weaker near-UV lines of Na\,{\sc i} at 3302~\AA, which are not included in the McDonald data.

HD~149757 lies close on the sky to the LDN 204 filamentary cloud complex and ionizes the H\,{\sc ii} region Sh~2-27. Radio 
observations of $^{12}$CO \citep{leb89,tac00,lis09}
indicate that the cloud is accelerated by the radiation from $\zeta$~Oph, causing compression of the molecular cloud and 
dense cores. This is supported by the CO observations of \cite{tac12}, who find several cloudlets of scales 
$\sim$1000 to 10,000 au which they suggest are cold, dense and overlapping. These structures could explain the apparent small-scale variations.

\subsection{HD~91316  Na\,{\sc i} D variations}

\begin{figure}
   \centering
   \includegraphics[width=\columnwidth]{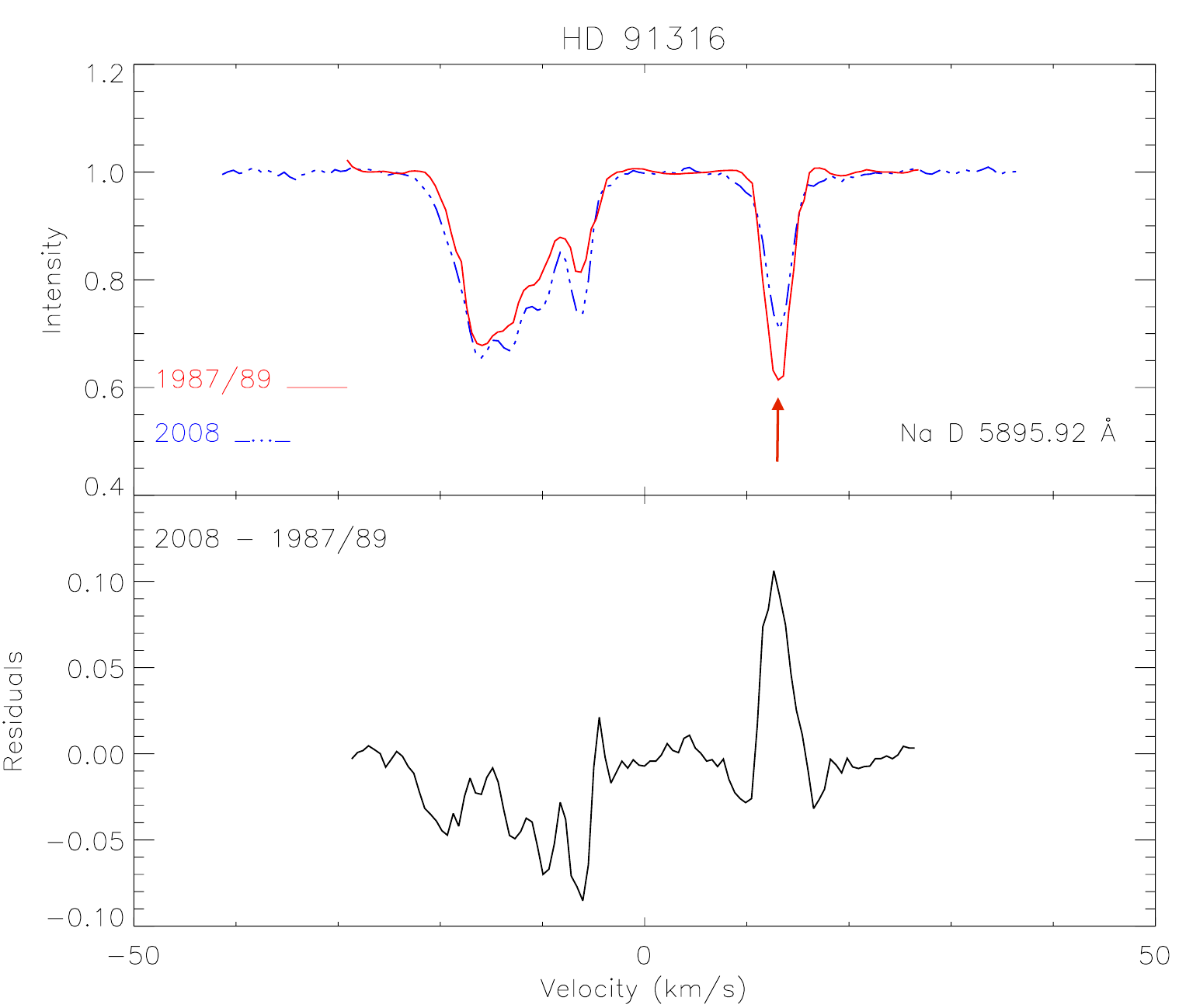}
   \caption{Overlay of WHK (solid red line) and McDonald spectra (dashed blue line line) for HD~91316 ($\rho$~Leo) showing variations at $23$\kms\ in the Na\,{\sc i} D line profile at 5895 \AA. The residuals between epochs are plotted below. It was necessary to degrade the resolution of the data used by WHK. } 
   \label{f_HD91316}
\end{figure}

HD 91316 ($\rho$~Leo) is a well-studied pulsating variable star at a distance of ~1350 pc, and is a frequent target for interstellar absorption-line studies. \citet{lau03} previously discovered variations in the Na\,{\sc i} D absorption-lines between 1995 and 2003 due to stellar motion of $\sim12$~au. Their epochs overlap those probed by our McDonald spectra, but used a different instrument. They found that the column density of Na\,{\sc i} at $v=18$\kms\ (Heliocentric velocity) decreased by $\approx20\%$ in 8 yr, but saw no change in Ca\,{\sc ii}. We confirm the variations in Na\,{\sc i} found by \citeauthor{lau03}, with the same component showing a similar variation. This is evident from (Fig. \ref{f_HD91316}) but, as was shown for HD 149757, the difference in resolution between the spectra has caused some problems in the analysis of this sightline. Due to the limited spectral range available in the WHK spectra there is not much continuum visible on either side of  the interstellar features and so normalisation would be difficult in this case. As noted by \cite{lau03}, the continuum may have been fitted slightly too low, near the varying component at 23 \kms, however adjusting for this would only yield even larger differences and so cannot be an explanation for this variation. The McDonald data are plotted in the figure along with the WHK spectrum, degraded in resolution to allow for a more accurate comparison. This variation is apparent in the residual plot and so this may be showing quite strong variation in the ISM over $\approx$ 20 yr. However, all components of this line vary between epochs. The most redward component varies in the opposite direction to the others (believed to be affected by differences in resolution ) which helps to confirm the variation, but makes quantifying it difficult.  This variation is over a scale up to 184~au, the transverse distance moved by the background star. \footnote{We are unable to reconcile the proper motion measured by van Leeuwen (2007) with the transverse distance of 12 au in 8 yr reported by \citeauthor{lau03}.}

\end{document}